\documentclass [12pt,preprint]{aastex}
\usepackage{epsfig}
\begin{document}
\voffset-1cm
\newcommand{\gsim}{\hbox{\rlap{$^>$}$_\sim$}}
\newcommand{\lsim}{\hbox{\rlap{$^<$}$_\sim$}}

\title{On the X-Ray emission of Gamma Ray Bursts}

\author{Shlomo Dado\altaffilmark{1}, Arnon Dar\altaffilmark{1}
and A. De  R\'ujula\altaffilmark{2}}

\altaffiltext{1}{dado@phep3.technion.ac.il, arnon@physics.technion.ac.il,
dar@cern.ch.\\
Physics Department and Space Research Institute, Technion, Haifa 32000,
Israel}
\altaffiltext{2}{alvaro.derujula@cern.ch; Theory Unit, CERN,
1211 Geneva 23, Switzerland \\ 
Physics Department, Boston University, USA}

\begin{abstract}

Recent data gathered and triggered by the SWIFT satellite have greatly 
improved our knowledge of long-duration gamma ray bursts (GRBs) and X-ray 
flashes (XRFs). This is particularly the case for the X-ray data at all 
times. We show that the entire X-ray observations  are in excellent 
agreement with the predictions of the `cannonball' model of GRBs and 
XRFs, which are based on simple physics
and were published long before the launch of SWIFT. Two 
mechanisms underlie these predictions: inverse Compton scattering and 
synchrotron radiation, generally
dominant at early and late times, respectively. The 
former mechanism provides a unified  description of the 
 gamma-ray peaks, X-ray flares and even the optical `humps'
seen in some favourable cases; i.e.~their very
different durations, fluxes and peak-times are related precisely
as predicted.
The observed smooth or bumpy fast decay of the X-ray light curve is 
correctly described case-by-case, in minute detail. The  
`canonical' X-ray plateau, as well as the subsequent
gradual steepening of the 
afterglow to an asymptotic power-law decay, are  as foretold.
So are the chromatic and achromatic properties of the light-curves.
\end{abstract}

\section{Introduction}

Since the launch of the SWIFT satellite, precise data from its Burst Alert 
Telescope (BAT) and X-Ray Telescope (XRT) have been obtained on the 
spectral and temporal behaviour of the X-ray emission of long-duration 
$\gamma$-ray bursts (GRBs) and X-ray flashes (XRFs)
from their beginning  until late times. 
The early
data are often complemented by the ultraviolet-optical telescope
(UVOT) on board SWIFT, and by ground-based robotic 
and conventional telescopes.
The ensemble of these
 data have already been used to test the most-studied theories of 
long duration GRBs and their afterglows (AGs), the 
{\it Fireball} (FB) models (see, e.g.~Zhang \& M\'esz\'aros~2004 and 
references therein) and the {\it Cannonball} (CB) model [see, e.g.~Dar \& 
De R\'ujula~2004 (hereafter DD2004); Dado, Dar \& De R\'ujula~2002a, 2003a
(hereafter DDD2002a, DDD2003a), and references therein]. 

The general behaviour of the SWIFT X-ray data has been described as 
`canonical' (Nousek et al.~2006; O'Brien et al.~2006; Zhang~2007). 
When measured early enough, the X-ray emission has prompt peaks that 
coincide with the $\gamma$-ray peaks of the GRB. The prompt X-ray flares 
have a continued fast temporal decline after the last detectable peak of 
the GRB. This rapid X-ray decline of the prompt emission ends within a 
couple of hundreds of seconds. Thereafter, it turns into a much flatter 
`plateau', typically lasting thousands to tens of thousands of seconds. 
Finally, the X-ray fluence, within a time of order one day, steepens into 
a power-law decline which lasts until the X-ray AG becomes too dim to be 
detected. Often, there are also X-ray peaks during the fast-decline phase 
or even later, not coinciding with a detectable $\gamma$-ray activity.

Neither the general trend, nor the frequently complex structure of the 
SWIFT X-ray data were predicted by (or can be easily accommodated within)  
the standard FB models (see, e.g.~Zhang \& M\'esz\'aros 2004, Piran~2005 
for reviews). Much earlier confrontations between predictions of the FB 
models and the observations also
 provided severe contradictions, such as the 
failure to understand the prompt spectrum on grounds of synchrotron 
radiation (e.g.~Ghisellini~2001) or the `energy crisis' in the comparison 
of the bolometric prompt and AG fluences (e.g.~Piran~1999, 2000). These 
contradictions have been disregarded. Many other problems of FB models 
have been commented upon (DD2004, Dar~2005 and references therein), 
including those related to `jet breaks' (e.g.~DDD2002a; Dar~2005; Dado, Dar 
\& De R\'ujula 2006), a-posteriori explanations of the reported detections 
of large $\gamma$-ray polarization (Dado, Dar \& De R\'ujula~2007b, and 
references therein) and radio AGs (e.g.~DDD2003a), 
and also ignored.

The SWIFT data have began to take their toll on the prevailing views on 
GRBs. Recently, Kumar et al.~(2007) concluded 
that `the prompt $\gamma$-ray emission 
cannot be produced in internal shocks, nor can it be produced in external 
shocks', and that `in a more general sense, $\gamma$--ray generation 
mechanisms based on shock physics have problems explaining the GRB data'. 
As for the X-ray AG, Burrows and Racusin (2007) examined the XRT 
light curves of the first $\sim\! 150$ SWIFT GRBs and reported that 
`although we expected to find jet breaks at typical times of 1-2 days 
after the GRB, we find that these appear to be extremely rare'. Curran et 
al.~(2006) have also carefully examined SWIFT data on GRB AGs and 
found that `X-ray and optical afterglows demonstrate achromatic breaks at 
about 1 day which differ significantly from the usual jet break in the 
blastwave model of afterglows'.

In spite of the above conundra, not all authors are so critical. Some posit
that the SWIFT data require only some modifications of the 
standard FB models in order to accommodate the results 
(e.g.~Panaitescu et al.~2006; Dai et al.~2007;  
Sato et al.~2007). Others still view the situation with faith: 
`Afterglows in the SWIFT era, contrary to expectations, did not allow us 
{\it to fully confirm yet} one of the most fundamental features of the 
standard afterglow picture: the presence of an achromatic break in the 
decaying light curve' (Covino et al.~2006, the emphasis is ours).

The situation concerning the CB model is  different. It successfully 
described the broad-band AGs observed before the SWIFT era (e.g.~DDD2002a; DDD2003a). 
This allowed us to establish the long-GRB/SN association, spectacularly
corroborated in the case of GRB 030329 (Dado, Dar \& De R\'ujula, 2003d; 
Stanek et al.~2003; Hjorth et al.~2003).
With use of the typical parameters and distributions extracted 
from the AG data, the model was used with success to predict all of the 
properties of the individual $\gamma$-ray peaks in the prompt 
emission of long GRBs (DD2004). It also correctly predicted the 
`canonical' trend of the X-ray emission observed by SWIFT, from the start 
of the initial fast decline, onwards (Dado et al.~2002a, 2006). The CB model was also used in foretelling the now well-established  
correlations between prompt ($\gamma$-ray) observables 
(Dar \& De R\'ujula~2000, DD2004, Dado, Dar \& De R\'ujula~2007b) 
and between prompt and AG observables (DD2004, Dado, Dar 
\& De R\'ujula~2007c). The model is summarized in 
$\S$\ref{CBMODEL}, its confrontation with data in $\S$\ref{outlook}.

We had previously posited and concluded (DD2004 and references therein, 
Dado et al.~2006) that three mechanisms successively dominate the 
radiation of a GRB: Compton scattering in the prompt $\gamma$-ray phase, 
thermal bremsstrahlung and line emission in the fast-declining X-ray 
phase, synchrotron radiation thereafter. The line emission phase was 
supported by the claimed observations of X-ray lines in early GRB 
afterglows (Piro et al.~1998;  Yoshida et al.~1999, 2001;  Piro et 
al.~2000; Antonelli et al.~2000; Reeves et al.~2002,
Watson et al.~2003)
and their very natural CB-model interpretation 
(Dado, Dar \& De R\'ujula 2003c). But these observations were of very 
limited statistical 
significance, and a phase during which line-emission significantly 
contributes may not, after all, be inevitably required.

In this paper we extend our analysis of X-ray emission in GRBs to include the prompt phase, during which $\gamma$-ray emission is abundant, and 
to the description of the flaring activity during the rapid decline of the 
X-ray signal. In this respect, and in our view, SWIFT has provided
two main observations on long GRBs and XRFs:  that an X-ray light curve, 
if  observed during the prompt phase, has a peak 
structure similar to that of the 
$\gamma$-ray light curve; that the generation of X-ray flares often continues 
much longer into the early AG phase. The
 observed prompt spectrum in the $\gamma$-ray to X-ray domain is that 
predicted for the prompt emission (DD2004), which in the CB model is 
Compton-dominated. The observed widths of the $\gamma$-ray 
and X-ray peaks, as well as lag-times between them
and their relative fluences, are in accordance 
with the model's predictions.  
This prompts us to investigate whether or not our model can be further
simplified, to describe the data in terms of only two mechanisms 
(Compton scattering and synchrotron radiation). We shall 
see that this simple picture, explicitly based on the predictions in 
DDD2002a and DD2004, gives a straight-forward and impressively successful 
description of the X-ray data, at all observed energies and times.

The SWIFT X-ray data show a flaring activity in a large fraction of GRBs.
The early X-ray peaks follow the pattern of the  
$\gamma$-ray pulses, they must have a common origin.
Let us refer to these $\gamma$-ray {\it and} X-ray peaks as `prompt'.
In the CB model,  
inverse Compton
scattering (ICS) is the origin of the prompt peaks, as we
review in $\S$\ref{Inverse}. Each peak is generated by a cannoball
emitted by the `engine', the accreting compact object resulting from a
core-collapse supernova (SN) event.
We shall see that ICS correctly describes the prompt peaks,
extending even into the optical domain in  cases in
which the relevant observations are available, such as in 
GRB060206 and XRF060218.

On occasion, superimposed on the declining X-ray afterglow, there are 
`late' X-ray flares, whose peak intensities decrease with time and whose 
accompanying $\gamma$-ray emission is below the detection sensitivity of 
BAT.
Yet, as we shall discuss, their spectral and temporal behaviour 
is similar to that of the prompt X/$\gamma$ pulses. Their natural
explanation is the same as that of the stronger flares: ICS
by the electrons of a CB. These CB emissions
must correspond to a weakening activity of the engine,
as the accreting material becomes scarcer.

In the CB model, from the onset of the `plateau' onwards, the X-ray,
optical (DDD2002a) and radio (DDD2003a) afterglows are in general dominated
by synchrotron radiation, the CB-model predictions for which
are reviewed in $\S$\ref{Synchrotron}. On occasion
these AGs also have transient rebrightenings (`very late' peaks),
one notable case being GRB030329 (Lipkin et al.~2004). During these 
episodes, the spectrum does not significantly change, it continues
to coincide with the one predicted on the basis of the 
synchrotron mechanism that dominates the late AGs. These very late peaks
are well described by encounters of CBs with density inhomogeneities
in the interstellar medium (DDD2002a, DD2004). Very late
peaks and bumps in the X-ray afterglow may have this origin as well.

We present in $\S$\ref{Case}
 a comparison between the CB model and the observations of 
a representative set of GRBs and XRFs with X-ray and optical light curves 
which are well sampled, have good statistics, and a relatively long follow 
up. As case studies for our analysis, we have chosen GRBs 060526, 050315, 
050319, 060729, 061121 and 060206. This sample includes the brightest of 
the SWIFT GRBs, the one with the longest measured X-ray emission, a couple 
with canonical X-ray light curves (with and without X-ray flares) and some 
of the allegedly most peculiar GRBs. In the CB model GRBs and XRFs are one 
and the same (DD2004; Dado et al.~2004) --the general distinction being 
that XRFs are viewed at a larger angle relative to the direction of the 
approaching jet of CBs. Thus we include two XRFs in our analysis: 050406 
and 060218. The second one is particularly interesting. Its optical AG, at 
various frequencies, shows not only the late peaks of the associated 
supernova, but a series of broad peaks at some 40000 s after 
trigger. We interpret these enhancements as the optical counterparts of 
the sole prompt X-ray peak of this XRF. 
The corresponding expressions describing  an ICS-generated
peak at all frequencies
(DD2004)  allow us to predict the positions and 
magnitudes of the optical enhancements, a gigantic extrapolation in time, 
radiated energy and 
frequency. The predictions are fulfilled, the interpretation is correct.

The CB model continues to be extremely successful in the confrontation
of its predictions with data.
This paper can be summarized in two sentences: Our description of
the observations is solely based on our published predictions; The figures
convey the quality of the results.

\section{The CB Model}
\label{CBMODEL}

In the CB model (e.g.~DD2004 and references therein)  {\it 
long-duration} GRBs and their AGs are produced by bipolar jets of CBs 
which are ejected (Shaviv \& Dar, 1995; Dar \& Plaga 1999)
in~{\it ordinary core-collapse} supernova explosions\footnote{Supernovae
associated with GRBs are viewed uncommonly
close to their CB jet axis, near which
the non-relativistic SN ejecta are faster than average. Their observed
large velocities may,
erroneously in our view, lead to their interpretation of the SNe as a special 
class:~{\it `hypernovae'}.}.
An accretion disk or a torus is 
hypothesized to be produced around the newly formed compact object, either 
by stellar material originally close to the surface of the imploding core 
and left behind by the explosion-generating outgoing shock, or by more 
distant stellar matter falling back after its passage (De R\'ujula~1987). 
As observed in microquasars (e.g.~Mirabel \& Rodriguez~1999; Rodriguez \& 
Mirabel~1999 and references therein), each time part of the accretion disk 
falls abruptly onto the compact object, a pair of CBs made of {\it 
ordinary-matter plasma} with a typical baryonic number, 
$N_{_{\rm B}}\!\sim\! 10^{50},$ are 
emitted with large bulk-motion Lorentz factors, typically 
$\gamma_0\!\sim\! 10^3$, 
in opposite directions along the rotation axis, wherefrom matter has 
already fallen back onto the compact object, due to lack of rotational 
support. 

The $\gamma$-rays of a single pulse of a GRB are produced as a CB coasts 
through the SN {\it glory} --the initial 
SN light, scattered by the `wind': the ejecta
puffed by the progenitor star in a succession of pre-SN flares. 
The electrons enclosed in the CB Compton up-scatter the photons of the 
glory to GRB energies. The initial fast expansion of the CBs and the 
increasing transparency of the wind environment 
 as the CBs penetrates it, result in the fast rise of GRB pulses. As 
the CB coasts further through the SN glory, the distribution of the glory's
light becomes increasingly radial and its density decreases rapidly. 
Consequently, the energy of the up-scattered photons is continuously 
shifted to lower energies and their number decreases 
rapidly. Typically, the ensuing fast decline 
of the prompt emission is taken over, 
within few minutes of observer's time,
by a broad-band afterglow --synchrotron emission from 
swept-in ISM electrons spiraling in the CB's enclosed magnetic field. The 
above picture leads to simple testable predictions for the intensity, 
spectrum and temporal evolution of the radiation emitted during the GRB 
phase (DD2004) and during the afterglow phase (DDD2002a; DDD2003a).
We summarize these predictions in $\S\S$ \ref{Inverse} and \ref{Synchrotron}.

\section{An interlude on nomenclature: `prompt' and `afterglow' radiations}

There is a good reason to put quotation marks in the title of this section.
Once upon a time, in pre-SWIFT days, the distinction was relatively clear.
More often than not, the `prompt' $\gamma$-rays defining a GRB were 
observed for a few seconds or minutes. The softer `afterglow' radiations 
were generally observed significantly later. 
SWIFT and robotic telescopes, satisfying their goal, 
have made this timing distinction obsolete: many GRBs have been observed
to shine also at `prompt' times at smaller energies than those of 
$\gamma$- or X-rays. Above their corresponding detectability thresholds,
a continuous `blending' of the observable radiations
at all frequencies and times has been observed. This is particularly true
for the X-ray monitoring on which SWIFT excels.

In the CB model, as in the observations, there is no clear-cut distinction
between prompt and {\it after}-glow signals. There are, however, two rather
distinct radiation mechanisms: inverse Compton scattering and 
synchrotron radiation (we say `rather' because synchrotron radiation
is but Compton scattering on virtual photons and because,
strictly speaking, in a universe whose
age is finite, all observed photons were virtual). For all cases we
have studied (by now, scores of GRBs and XRFs), the
$\gamma$ and X-ray radiations are dominated by ICS
until the end of their fast declining phase. Synchrotron radiation dominates
the X-ray production from its `plateau' onwards. All other
general statements equating the two mechanisms to the two phases
(prompt and afterglow)
have known exceptions. Two examples:

There are cases, such as GRB 061126, in which
the observed UVOIR emission (Perley, et al.~2007) 
during the prompt $\gamma$-ray
 phase is synchrotron dominated.  Since we do not study any such
 case here, suffice it to explain the reason. After correcting for 
extinction, the spectral energy density of the synchrotron 
emission increases with decreasing frequency. Contrariwise,
the ICS spectrum,
below its peak energy or exponential cutoff,
 is approximately flat. Consequently synchrotron emission can 
contribute significantly, and even dominate the observed 
(lower-frequency) UVOIR emission during the `prompt' phase. 
We shall demonstrate that there are 
also cases, such as XRF060218, in which ICS is a significant 
contribution to the optical afterglow until the time it becomes dominated
by the associated SN. 

In spite of the above, we shall not always be over-pedantically precise
and, when there is no room for confusion, we shall often refer to 
the prompt phase as synonymous with ICS dominance and to the
afterglow phase as tantamount to synchrotron dominance.

\section{Inverse Compton Scattering}
\label{Inverse}

\subsection{The spectrum of GRB pulses}

During the initial phase of $\gamma$-ray emission in a GRB,
the Lorentz factor $\gamma$ of a CB stays put at its initial value
$\gamma_0\!=\!{\cal{O}}(10^3)$, for the deceleration induced by 
the collisions with the ISM has not yet had a significant effect
(DD2000b, DDD2002a, DDD2003a).
Let $\theta$ be the observer's angle relative to the direction of motion
of a CB.
The Doppler factor by which light emitted by a CB is boosted in energy
is 
\begin{equation}
\delta={1\over \gamma\,
(1-\beta\, \cos\theta)}\approx {2\, \gamma\over
(1+\gamma^2\, \theta^2)}\; ,
\label{delta}
\end{equation}
 where the 
 approximation is excellent for $\gamma\gg 1$ and $\theta\ll 1$.
The emitted light is forward-collimated into a cone of 
characteristic opening angle $1/\gamma$,
so that the boosted energetic radiation is observable for 
$\theta\!=\!{\cal{O}}(1/\gamma_0)$. This implies that the typical
initial Doppler factor of a GRB is
$\delta_0\!=\!{\cal{O}}(10^3)$.

The SN glory has a thin thermal-bremsstrahlung spectrum: 
\begin{equation}
\epsilon\, {dn_\gamma \over d\epsilon} \sim \left({\epsilon \over 
\epsilon_g }\right)^{1-\alpha}\, e^{-\epsilon/\epsilon_g}, 
\label{thinbrem}
\end{equation}
with a typical (pseudo)-temperature,
$ \epsilon_g \!\sim\!1$ eV, and index $\alpha\!\sim\!1$.
The observed energy of a glory's photon, inverse
Compton scattered by an electron comoving with a 
CB at redshift $z$, is:
\begin{equation}
E={\gamma_0\, \delta_0\, \epsilon \, (1+\cos\theta_i)\over (1+z)}   
\label{ICSboost}
\end{equation}
where $\theta_i$ is the angle of incidence of the initial
photon onto the CB, in the SN rest system.

The predicted GRB spectrum is (DD2004): 
\begin{equation}
E\, {dN\over dE} \sim \left({E\over T}\right)^{1-\alpha}\,
 e^{-\epsilon/T}+ b\,(1-e^{-E/T})\, \left({E \over T}\right)^{-p/2} .
\label{GRBspec}
\end{equation}
The first term, with $\alpha\!\sim\! 1$, is the result of Compton 
scattering by the bulk of the CB's electrons, which are comoving with it.
The second term in Eq.~(\ref{GRBspec}) is induced by 
a very small fraction of
`knocked on' and Fermi accelerated electrons, whose initial spectrum
(before Compton and synchrotron cooling) is $dN/dE_e\propto E_e^{-p}$, 
with $p\approx 2.2$. Finally, $T$ is the effective (pseudo)-temperature 
of the GRB's photons:
\begin{equation}
T\equiv{4\, \gamma\, \delta\,\epsilon_g\,
\langle 1+\cos\theta_i\rangle \over 3\, (1+z)}
\approx (356 \,{\rm keV})\;{\gamma_0\,\delta_0\over 10^6}\;
{\epsilon_g\over 1\,{\rm eV}}\;{3.75\over 1+z}\;
\langle{1+\cos\theta_i}\rangle,
\label{ICST}
\end{equation}
where, in the numerical result, we have normalized to typical values,
including the mean redshift $\langle z\rangle = 2.75$ of SWIFT's
long GRBs.
For a semi-transparent glory $\langle\cos\theta_i\rangle$ would be
somewhat smaller than zero.

For $b={\cal{O}}(1)$,
the energy spectrum predicted by the CB model, Eq.~(\ref{GRBspec}),
 bears a striking resemblance 
to the Band function (Band et al.~1993) traditionally used to model the 
energy spectra of GRBs. For many SWIFT GRBs the spectral observations
do not extend to energies much bigger than $T$, or the value of $b$
in Eq.~(\ref{GRBspec}) is relatively small, so that the first term 
of the equation provides a very good approximation.
This term coincides with the `cut-off 
power-law' spectrum which has also been recently used to model 
many GRB spectra. It yields a peak value of $E^2\, dN/dE$ at 
$E_p\!=\!(2\!-\!\alpha)\, T\!\approx\! T$ for 
$\alpha\!\sim \!1$  (for $b\!=\!\alpha\!=\!1$, $E_p\!\approx\! 1.5\, T$).
The predicted range of $E_p$ values, 
for $T$ as in Eq.~(\ref{ICST}), is in good agreement with the 
observations of BATSE, 
BeppoSAX, Konus-Wind, Integral and RHESSI, which cover a much broader 
energy range than SWIFT\footnote{SWIFT data can determine
 $E_p$ only when it is 
well within its 15-350 detection range. This results in a biased sample 
of GRBs whose (measured) $E_p$ is smaller than the average 
over the entire GRB population.}.

In the CB model XRFs are the same objects as GRBs, the main difference
being that the typical observer's angle is larger in the former than in the
latter (DD2004, Dado, Dar \& De R\'ujula 2004a). 
For similar values of $\gamma_0$, this implies a smaller $\delta_0$
in Eq.~(\ref{delta}),
and consequently the softer spectrum and
relatively small $E_p$ that define an XRF,
see Eqs.~(\ref{ICSboost}-\ref{ICST}). XRFs have lightcurves
with wider and less rugged peaks than GRBs. This follows from
the time dependence of the corresponding light-curves, which
we discuss next.

\subsection{The lightcurves of GRB and XRF pulses in various energy
ranges}

Two `transparency times' and a  `cooling time'
play a role in the determination of the
shape of the lightcurve of a pulse in a GRB or an XRF (DD2004). 
For typical parameters, they happen to be of very similar order
of magnitude, which simplifies the study of the incidence of their effects.
The first characteristic 
time is the one it takes the plasma within a CB to become 
transparent to the radiation the observer sees as $\gamma$- or X-rays.
Let $\sigma_T=0.665\times 10^{-24}$ cm$^{-2}$ be the Thomson cross 
section.
A CB of approximately equal electron and baryon number $N_{_{\rm B}}$,
initially expanding in its rest system
at a speed $\beta_s$ (in units of the speed of
sound in a relativistic plasma, $c/\sqrt{3})$, becomes transparent
when its radius reaches a value 
$R_{tr}\!\sim\!\sqrt{3\, \sigma_T\, N_{_{\rm B}}/(4\, \pi)}\!\approx \! 
(4\times 10^{12}\,{\rm cm})\, \sqrt{ N_{_{\rm B}}/ 10^{50}}$. 
Seen by a cosmological observer, the time elapsed from the CB's
ejection to the transparency time is:
\begin{equation}
t_{tr}=(0.86\,{\rm{s}})\; {1\over \beta_s}\;{10^3\over\delta}\;
{1+z\over 3.75}\;\left({N_{_{\rm B}}\over 10^{50}}\right)^{1\over 2}.
\label{ttrans}
\end{equation}

The typical wind-fed surroundings of a SN have a course-grained
density distribution $\rho\!\propto\! 1/r^2$, typically such that
$ \rho \,r^2 \! \sim \! 10^{16}\, {\rm g\, cm^{-1}}$. 
The probability that a GRB photon produced at a distance $r$ from the SN
to evade being absorbed
is ${\rm exp}[-r^w_{tr}/r]$ with
$r^w_{tr}=\sigma_T\, \rho\, r^2/m_p$  the distance at which the 
remaining optical
depth of the wind is unity. In the SN rest frame the
`wind transparency time'  is $r^w_{tr}/c$, and it corresponds
to an observer's time:
\begin{equation}
t^w_{tr}=(0.51\,{\rm s})\;{\rho\,r^2\over 10^{16}\,{\rm g \,cm}^{-1}}\;
{1+z\over 3.75}\; {10^6\over \gamma\,\delta}\; ,
\label{twtrans}
\end{equation}
coincidentally close ---for the typical parameters---
to the CB's transparency time, $t_{tr}$. 
Our characterization of `typical' wind densities and profiles is
rough. There should be significant case-by-case deviations
from the reference time of Eq.~(\ref{twtrans}).

In the inner non-transparent regions of a wind, 
$\langle\cos\theta_i\rangle\!\simeq\! 0$ in Eq.~(\ref{ICSboost}), 
but no photons come out. As the wind becomes transparent,
its scattered photon population decreases as 
$1/r^2\!\propto\! 1/t^2$ and so does the number of Compton
up-scattered photons, on a time scale of order $t^w_{tr}$. On the
same time scale, the glory's photons become more radially directed,
so that $\langle 1+\cos\theta_i\rangle\to t^w_{tr}/t^2$. 
According to Eq.~(\ref{ICSboost}), this implies that their typical
energies also decrease as $1/t^2$. This means that, as the
wind's transparency sets in exponentially fast, 
a pulse's energy fluence $E\,d^2N_\gamma/dEdt$ becomes,
equally fast, an approximate
function of the combination $E\,t^2$ of its two variables.
Finally, the index $p$ in Eq.~(\ref{GRBspec}) continuously 
varies from $p\!\simeq\!2.2$
at the start of a pulse to $p\!\simeq\!3.2$ towards its end.
This is because the accelerated  and knocked-on electrons within
a CB lose energy as a result of Compton 
and synchrotron cooling, on the same
time scale as a CB's Compton transparency time $t_{tr}$.

In DD2004 we have studied individually all of the above effects,
and the incidence of various hypothetical CB geometries.
Different CB shapes make little difference. For all of them
(except a slab-shaped CB seen at $\theta\!=\!0$) 
 the photons simultaneously arriving to an observer 
 have been Compton scattered at different depths
and times within a CB. This implies that the time-dependent effects
we have discussed are smoothed over a pulse's duration. We have seen,
for instance, why as the wind becomes transparent,
the energy fluence $E\,d^2N_\gamma/dEdt$ 
effectively becomes a function of $E\,t^2$. The geometrical averaging
over a CB's volume makes this result a good approximation at all 
times.

The initial rise of a pulse is exponentially fast,  
a result of
the exponential increase with time of a CB's transparent skin-depth, as
it expands prior to its transparency time, or of the wind's decreasing
remaining column density, as a CB travels through it.
The tail of a pulse behaves approximately as $1/t^2$,
reflecting the glory's photon number-density profile.

All of the above effects can be summarized in a {\it master formula}
for a pulse's energy fluence as a function of $E$ and $t$, a very good
approximation to all of the above effects, studied in more detail in DD2004.
To specify the master formula, let us first define an energy-dependent width
of a pulse:
\begin{equation}
\Delta t(E) \approx t_{tr}^w \left(T\over E\right)^{1\over 2},
\label{Deltat}
\end{equation}
where $T$ is given by
Eq.~(\ref{ICST}) and $t_{tr}^w$ is as in Eq.~(\ref{twtrans}). The choice
of this time scale, as opposed to $t_{tr}$ in Eq.~(\ref{ttrans}), is
dictated by the fact that in our study of correlations between GRB 
observables (Dado, Dar \& De R\'ujula 2007a) we have learned that, more often than not,
the wind transparency time plays the dominant role in specifying a
pulse's time properties (rise-time, lag-time and variability).
In this paper we are mainly dealing with high-quality X-ray data.
We consequently fit $\Delta t(E)$ in Eq.~(\ref{Deltat}) case-by-case 
and pulse-by-pulse to the
data at a particular X-ray energy band, test its explicit $E$-dependence
in other X-bands and, when the data is available, at the UVOT 
energies\footnote{In some cases the optical data are superior,
and we do the contrary. 
The $\gamma$-ray data are not analized well enough 
to determine precisely
the width of its pulses and the corresponding uncertainty.
We have checked that they are compatible in every case
with the prediction from the observed width of the sister X-ray
pulse and the energy-dependence of Eq.~(\ref{Deltat}).}.

Denote by an index `{\it i}' the i-th pulse in a GRB, produced by  
a CB launched at (an observer's) time $t_i$.
In terms of the pulse's width, $\Delta t(E)$, in a given observed energy range
$E$, we can define a time-dependent (pseudo)-temperature for the pulse:
\begin{equation}
T_i\approx {4\, \gamma_i \,\delta_i\, \epsilon_g \over 3\, (1+z) }\,
\left[1-{t-t_i\over \sqrt{\Delta t_i^2+(t-t_i)^2}}\right]\, ,
\label{ICSTi}
\end{equation}
where we allowed for different values of 
$\gamma$ and $\delta$ for different pulses in the same GRB;
even the emission angle may change, e.g.~due to precession 
of the ejection axis. 
The `master formula' for the energy fluence of a pulse as a function
of time and energy is: 
\begin{equation}
E\, {d^2N_\gamma[i]\over dt\,dE} \propto
\Theta[t-t_i]\;
e^{-[\Delta t_i/(t-t_i)]^m}\, 
 \left\{1-e^{-[\Delta t_i/(t-t_i)]^n}\right\}\, E\,{dN_\gamma[i] \over dE}  
\label{GRBlc}
\end{equation}
where  $m\approx n\approx 2$,
and $E\, dN_\gamma[i]/dE$ is given by 
Eq.~(\ref{GRBspec}) with $T\rightarrow T_i$. 

At X-ray energies and, more so, at smaller ones, 
the first term on the RHS of
Eq.~(\ref{GRBspec}) usually dominates  $E\, dN_\gamma/dE$. Consequently,
the light curve generated by a  sum of pulses
is generally well approximated by:
\begin{equation}
E\, {d^2N_\gamma\over dt\,dE}\approx 
\sum_i\,A_i\, \Theta[t-t_i]\;
e^{-[\Delta t_i/(t-t_i)]^2}\, 
 \left\{1-e^{-[\Delta t_i/(t-t_i)]^2}\right\}\, e^{-E/T_i},
\label{GRBXlc} 
\end{equation}
until the
ICS emission is overtaken by the broad band
synchrotron emission from the swept-in ISM electrons.
This is the expression to be used in our fits.
Notice that the individual peaks in Eq.~(\ref{GRBXlc}) obey
\begin{equation}
E\, {d^2N_\gamma\over dt\,dE}(E,t)\approx F(E\,t^2),
\label{law}
\end{equation}
to which we shall refer as the `$E\!\times\!t^2$~{\it law'}.
A simple consequence of this law is that the ICS peaks have that
same shape at all frequencies, scaled in time as $t\propto E^{-1/2}$.

The time dependence of $T_i$ in Eq.~(\ref{ICSTi}) results in a 
decreasing peak
energy of individual pulses as a function of time:
\begin{equation}
E_{p}^i(t)\approx 
E_{p}^i(t_i)\,
\left[1-{t-t_i\over \sqrt{\Delta t_i^2+(t-t_i)^2}}\right]\, .
\label{ept}
\end{equation}
For late times, $E_{p}^i$  has the approximate form
$E_{p}^i(t)\approx E_{p}^i(t_i)\,\Delta t_i^2/2\, (t-t_i)^2.$
Such an evolution
has been observed in the time resolved spectra of well isolated pulses 
(see, for instance, the insert in Fig.~8 of Mangano et al.~2007).

Three other trivial but important 
consequences of Eqs.~(\ref{Deltat}-\ref{law}) 
are the following (DD2004). The widths of a given pulse in
two energy bands (a $\gamma$-ray and an X-ray one, for instance)
are approximately related by:
\begin{equation}
\Delta t(E_1)\approx\Delta t(E_2)\left(E_2\over E_1\right)^{1\over 2}
\label{widthrelation}
\end{equation}
The onset-time, $t_i$, of a GRB pulse is simultaneous at all energies.
But the peak times, $t_p^i(E)$, differ (the lower-energy ones `lag'), 
and are approximately related by:
\begin{equation}
t_p^i(E_1)-t_i\approx 1.2\, \Delta t_i(E_1)
\approx (E_2/E_1)^{0.5}\, 
[t_p^i(E_2)-t_i]\, ,
\label{peaktimes}
\end{equation}
where the numerical factor 1.2 results from the explicit shape of the light curve in Eq.~(\ref{GRBXlc}). The full width at half maximum, again for this
pulse shape, is
\begin{equation}
{\rm FWHM}\!\approx\! 1.8\, \Delta t_i,
\label{fwhm}
\end{equation}
and it extends from 
$t\!\approx\! t_i\!+\!0.72\,\Delta t_i$ to $t\!\approx\!t_i\!+\! 2.5\, \Delta t_i$.
In Eqs.~(\ref{widthrelation},\ref{peaktimes},\ref{fwhm}) 
most of the case-by-case
variability factors in Eqs.~(\ref{ttrans},\ref{twtrans},\ref{ICSTi}) factor out.
They should consequently be rather good approximations.

\subsection{Optical `humps', X-ray `flares' and $\gamma$-ray pulses}

In more than 50\% of the GRBs observed by SWIFT, the X-ray light curve, 
during the prompt GRB and its early AG phase, 
shows flares superimposed on a smooth background. Some 
examples are shown in Fig.~\ref{f3}, borrowed from Nava et al.~(2007),
to which we have added the corresponding (linear time-scale) GRB light
curves. In these figures, the smooth `background' to the flares
 was decomposed by Willingale et al.~(2006) 
and by Nava et al.~(2007) into two separate contributions from the GRB
 and its AG, using an arbitrary parametrization. As in the
examples in Fig.~\ref{f3}, the $\gamma$-ray and X-ray light curves of 
GRBs show 
that most of the observed early X-ray flares are part of the emission in the
prompt GRB pulses.
This is the case for the entire sample of X-ray light curves fitted 
by Willingale et al.~(2006). Their distinction between
flares and `background', as well as the interpretation of the latter,
appear to us not to be justified.

In the CB model, an X-ray `flare' coincident in time with a $\gamma$-ray
pulse is simply its low-energy tail. They are both
due to Compton scattering of photons 
in the thin-bremsstrahlung spectrum of the SN glory. The glory's
photons incident on the CB at small $E_i$ or $1\!+\!\cos\theta_i$
result in X-ray or softer up-scattered energies, see Eq.~(\ref{ICSboost}).
The harder and less collinear photons result in $\gamma$-rays.
The light-curve and spectral evolution of an X-ray flare
are given by Eqs.~(\ref{ICSTi},\ref{GRBlc}).
Its width is related to that of the accompanying $\gamma$-ray pulse  
as in  Eq.~(\ref{widthrelation}). Relative to its $\gamma$-ray
counterpart, an X-ray flare is wider and  its peak time `lags',
see Eqs.~(\ref{peaktimes},\ref{fwhm}).
The X-ray flares during a GRB are well 
separated only if the $\gamma$-pulses are sufficiently spaced. 

Equations (\ref{ICSTi},\ref{GRBlc}) also describe the optical counterpart
of a flare. 
In agreement with the prediction of Eq.~(\ref{widthrelation}),
$\Delta t({\rm eV})\approx 10^3\, 
\Delta t({\rm MeV})$, optical flares during the prompt GRB are usually 
very wide and completely blended. 
They are well separated and observable as wide optical `humps'
in very long GRBs with 
very large spacing between narrow $\gamma$-ray pulses,
 such as GRB 060206, which we study below in detail. Another
very clear case which we analize is XRF 060218, 
with only one X-ray flare, whose optical counterparts are also 
 clearly visible as humps in the lightcurves at different wave-lengths.

In the CB model, X-ray flares without an accompanying
detectable $\gamma$-ray emission can be produced  by CBs 
ejected with relatively small  Lorentz factors
and/or relatively large viewing angles  (Dado et al.~2004).
Such CBs may be ejected in accretion episodes both 
during the prompt GRB and the subsequent times. 
The spectrum of these flares, Eq.~(\ref{GRBXlc}), is different 
from the synchrotron-emission spectrum,
$F_\nu\!\sim\! \nu^{-1.1}$, of late-time flares that result from 
the passage of CBs through local 
over-densities in the interstellar medium (DDD2003a).
We shall see in our case studies that very often, during the rapidly
decreasing phase of the X-ray light curve, there are `mini-flares'
not seen in the corresponding $\gamma$-ray light curve.
In the CB model, though we have not predicted it,
 this is perfectly natural. The CBs are ejected
 in delayed accretion episodes (De R\'ujula 1997) of matter from
a ring or torus (Dar and De R\'ujula, 2000a). As the accretion
material is consumed, one may expect  the `engine' to have
a few progressively-weakening dying pangs.

\section{Synchrotron radiation}
\label{Synchrotron} 

A second mechanism, besides ICS, generates radiation from a CB.
A CB encounters matter in its voyage through the 
interstellar medium (ISM), effectively
ionized by the high-energy radiation of the very same CB.
This continuos collision with the medium decelerates the CB in
a characteristic fashion, and results in a gradual steepening of the
light curves, which is achromatic if synchrotron radiation dominates
(DDD2002a). In $\S$\ref{Deceleration},
we review the calculation of $\gamma(t)$, the CB's diminishing
Lorentz factor. We have assumed and tested observationally, 
via its CB-model consequences,
that the impinging ISM generates within
the CB a turbulent magnetic field\footnote{`First principle' numerical
simulations of two plasmas merging at a relative relativistic
Lorentz factor (Frederiksen
et al.~2003, 2004; Nishikawa et al.~2003)
do not generate the desired shocks, but do generate turbulently
moving magnetic fields.}, in approximate energy equipartition
with the energy of the intercepted ISM (DDD2002a, DDD2003a). 
In this field, the intercepted
electrons emit synchrotron radiation. This radiation, isotropic in the 
CB's rest frame,
is Doppler boosted and collimated around the direction of motion 
into a cone of characteristic opening angle $\theta(t)\sim 1/\gamma(t)$. 
In $\S$\ref{SyncSpec} we summarize the predictions of the synchrotron 
radiation's dependence on time and frequency
 (DDD2002a, DDD2003a).

\subsection{The deceleration of a CB}
\label{Deceleration}

As it ploughs through the ionized ISM, a CB 
gathers and scatters its constituent ions, mainly protons. These encounters
 are `collisionless' since, at about the time it becomes transparent to 
radiation, a CB also becomes `transparent' to hadronic interactions
(DD2004). The scattered and re-emitted 
protons exert an inward pressure on the CB, countering its expansion.
In the approximation of isotropic re-emission in the CB's 
rest frame and a constant ISM density $n\!\sim\!n_e\!\sim\!n_p$, 
one finds that within minutes of observer's time $t$, a CB
reaches an approximately constant `coasting'
asymptotic radius $R\!\sim\!10^{14}$ cm, 
before it finally stops
and blows up, after a journey of years of observer's
time. During the coasting phase,
and in a constant density ISM, $\gamma(t)$ 
obeys (DDD2002a, Dado et al.~2006): 
\begin{equation}
({\gamma_0/ \gamma})^{3+\kappa}+
(3-\kappa)\,\theta^2\,\gamma_0^2\,(\gamma_0/\gamma)^{1+\kappa} 
= 1+(3-\kappa)\,\theta^2\,\gamma_0^2+t/t_0\,,
\label{decel}
\end{equation}
where
\begin{eqnarray}
t_0&=&{(1+z)\, N_{_{\rm B}}\over
(6+2\kappa)\,c\, n\,\pi\, R^2\, \gamma_0^3}
\nonumber\\
&\approx&
 (1.3\times 10^3\, {\rm s})\, (1+z)
\left[{\gamma_0\over 10^3}\right]^{-3}\,
\left[{n\over 10^{-2}\, {\rm cm}^{-3}}\right]^{-1}\,
\left[{R\over 10^{14}\,{\rm cm}}\right]^{-2}\,
\left[{N_{_{\rm B}}\over 10^{50}}\right],
\label{break}
\end{eqnarray}
where the numerical value is for $\kappa\!=\!1$, the result for the
case in which the ISM particles re-emitted fast by the
CB are a small fraction of the flux of the intercepted ones. In the
opposite limit, $\kappa\!=\!0$.
In all of our fits we use $\kappa\!=\!1$, though the results are
not decisively sensitive to this choice. In the CB model of cosmic
rays (Dar \& De R\'ujula 2006) the observed spectrum strongly
favours $\kappa\!=\!1$. For both limits of $\kappa$, $\gamma$  and $\delta$  
change little as long as $t\!<\!t_0$ and later tend to an asymptotic
power law, $t^{-1/(3+\kappa)}$. 
 This transition induces a gradually steepening 
{\it deceleration bend}  in the synchrotron AG
of a CB at $t \!\sim \! t_0$,
 achromatic all the way from X-ray to infrared
frequencies (DDD2002a, Dado et al.~2007a).

\subsection{The Synchrotron spectral energy density}
\label{SyncSpec} 

Let a prime denote a variable in a CB's rest frame.
A frequency $\nu'$, and time interval $dt'$ in the CB's rest frame 
are related to the frequency $\nu$ and arrival time interval $dt$ 
in the observer's frame 
through  $\nu'= (1+z)\,\nu/\delta$ and $dt'=\delta\,dt/(1+z)$. 
The energy flux density in the CB frame, generated by synchrotron 
radiation from the ISM electrons 
that enter the CB and radiate there the bulk of their incident energy,
is given by (DDD2002a, DDD2003a)
\begin{equation}
F_{_{\rm CB}}[\nu',t']\simeq \eta\,\pi\, R^2\,n_e\, m_e\ 
c^3\,\gamma(t)^3\,f_{\rm sync}(\nu',t')
\label{CBEFD}
\end{equation}
where $\eta\!\sim\! 1$ is the fraction of intercepted ISM 
electrons that enter the CB, 
and $f_{\rm sync}$ is the normalized spectral shape.
To specify $f_{\rm sync}$, let  $\nu'_b(t')$, the `bend frequency'
of the CB model (not to be confused with any
of the `break' frequencies of fireball models) 
be the characteristic synchrotron 
frequency radiated by the ISM electrons as they enter a CB at time $t'$ 
with a relative Lorentz factor $\gamma(t')$. 
In the observer's and CB's frames, the bend frequencies are:
\begin{equation}
\nu_b(t)={\delta\over 1+z}\;\nu_b'(t')
\simeq {5.91\times 10^{15} \over 1+z}\, 
{[\gamma(t)]^3\, \delta(t)\over 10^{12}}\,
\left[{n_p\over 10^{-2}\;\rm cm^3}\right]^{1/2}
{\rm Hz}.
\label{nub}
\end{equation}
Above the observer's radio frequencies,
at which self-absorption within the CB
and other effects are relevant (DDD2003a), 
$f_{\rm sync}$
 has the normalized shape:
\begin{equation}
f_{\rm sync}(\nu',t')={K(p)\over\nu'_b(t')}\, {[\nu'/\nu'_b(t')]^{-1/2}
                       \over\sqrt{1+[\nu'/\nu'_b(t')]^{p-1}}}\,,
\label{fsyn}
\end{equation}
where $p\!\approx\! 2.2$ is the the assumed spectral index of electrons,
prior to synchrotron cooling, that are 
Fermi-accelerated within the CB\footnote{In the CB model only a very
small fraction of electrons, the ones contributing to the spectrum of
Eq.~(\ref{fsyn}) at $\nu\ge\nu_b$, need be Fermi accelerated. The required acceleration
efficiency is correspondingly tiny.}, and the normalization factor is
\begin{equation}
K(p) \equiv \left\{
{2\over \sqrt{\pi}}\, \Gamma \left[ {2\,p-1\over 2\,(p-1)} \right]\,
\Gamma\left[{p-2\over 2\,(p-1)}\right] 
\right\}^{-1}
\approx {p-2 \over 2\,(p-1)}\,.
\label{K}
\end{equation}

The spectral energy 
density of the synchrotron radiation
from a single CB at a luminosity distance $D_L$, as
seen in the observer's frame, is given by 
\begin{equation}
F_{\rm obs}[\nu, t] \simeq {(1+z)\,\delta(t)^3\over 4\,\pi\, D_L^2}\, A(\nu,t)
F_{_{\rm CB}}[\nu',t'] 
\label{Fnu}
\end{equation}
where 
$A(\nu, t)$ is the attenuation along the line of sight.
In the CB model the attenuation in the host galaxy may be a function of
time, for the CB's travel long distances within it.

\section{Chromatic and achromatic light curves}
\label{Chromo}

Inverse Compton scattering on the glory's light
results in a very specific chromatic behaviour,
summarized by the dependence of the
 energy fluence of Eq.~(\ref{law}) on the combination
of variables $E\!\times\!t^2$. At times and frequencies at which ICS
contributes significantly, the light curves have the consequent
 chromatic behaviour, a fact predicted and checked
in DD2004 for the $\gamma$-rays of GRBs, to be corroborated in 
detail at lower energies in $\S$\ref{Case}. 

There are cases in which
ICS contributes significantly to the optical bands at sufficiently
late times for the X-rays to have already reached their plateau
synchrotron-dominated AG. This results in a very specific 
chromatic behaviour: the X-ray lightcurve is smooth, while the
optical lightcurves `bend-up', or even display humps that
are nothing but the X-ray pulse(s) delayed and widened by the
$E\!\times\!t^2$ `law'. A striking example of this very
chromatic behaviour is shown in
Fig.~(\ref{f1}), discussed in more detail in $\S$\ref{Case}. 

The colour properties of synchrotron radiation are complex,
even in what is by far the most common case: that the optical
`afterglow' is  synchrotron-dominated. The simplest situation 
arises when all observed frequencies are below the injection bend; 
notice that for typical reference parameters,
$\nu_b(0)$ in Eq.~(\ref{nub}) corresponds to an energy of 
${\cal{O}}(10\,\rm eV)$. If $\nu_b(t)$ at the time of the observations
is below the observed bands,
the (unattenuated) synchrotron spectrum stays 
put at $F_\nu\!\sim\!\nu^{-1.1}$, see Eq.~(\ref{fsyn}), and is
achromatic all the way to X-rays, if the light curve of the latter is observed
after its rapidly decreasing, ICS-dominated phase. Many such
cases have been studied, e.g.,~in DDD2003a. 

The achromatic behaviour we just discussed is particularly striking
when the X-ray and optical light curves are observed to change
fast the trend of their
time dependence  at the end of their `plateau' phase.
In the CB model this occurs at the `deceleration bend' $t_0$ of
Eq.~(\ref{break}). Around that time, according to 
Eqs.~(\ref{break},\ref{Fnu}), $F_{\rm obs}[\nu,t]$ steepens to
a power-law decline, 
$F_{\rm obs}[\nu,t]\!\propto\! 
\nu^{-p/2}\,t^{-(p+1)/2}\!\approx\! \nu^{-1.1}\,t^{-1.6}$
for $\kappa=1$, or $\propto \!
\nu^{-p/2}\,t^{2(p+1)/3}\!\approx\! \nu^{-1.1}\,t^{-2.13}$
for $\kappa=0$. 
This smooth CB-deceleration `bend' is not to be confused 
with the unobserved 
achromatic {\it break}  predicted in fireball models 
(Rhoads 1998). The CB model interpretation of this well understood
achromatic bend 
(see, e.g.~DDD2003a) is further strengthened by the fact that it
is observed at the predicted time scales, which display the
predicted correlations with the time-integrated
isotropic-equivalent energies
of the GRB's prompt X-rays and of the plateau X-ray AG (Dado et al.~2007c).

The next simplest
situation arises when $\nu_b(0)$ is above the optical bands
and the data start early enough for the decreasing $\nu_b(t)$ to be 
observed as it `crosses' the optical frequencies. In that case the 
(unabsorbed) optical
spectrum evolves in a predicted fashion from $F_\nu\!\sim\!\nu^{-0.5}$
to $F_\nu\!\sim\!\nu^{-1.1}$, while the (unabsorbed) X-ray flux stays
put at the $F_\nu\!\sim\!\nu^{-1.1}$ tail of the distribution, see 
Eq.~(\ref{fsyn}). Many cases of this very specific chromatic
evolution have been studied in DDD2003a. The success of their
CB-model description corroborates the assumption that the CB's
inner magnetic-field intensity $B(t)$, of which $\nu_b(t)$ is a
function, is approximately determined by the equipartition hypothesis.
 
A variation of the chromatic behaviour arising as a consequence of
bend-frequency crossing occurs when it happens early enough for
the circumburst density profile to be still dominated by the progenitor's
pre-SN wind emissions. 
At early times, $t\!\ll\! t_0$,
 the deceleration of CBs has not significantly affected their motion
and $\gamma(t)$ and $\delta(t)$ are practically constant.
Yet, the observer's
bend frequency, $\nu_b(t)\!\propto\! \sqrt{n(t)}\, [\gamma(t)]^3 \delta(t)$,
may decrease with time as $n(t)$ varies.
Keeping track of the $n$-dependence, we conclude (DDD2003a) from 
Eqs.~({\ref{CBEFD}-\ref{Fnu}) that the
(unextinct) early synchrotron radiation of a CB moving in a windy
density profile, $n\!\sim\! r^{-2}\!\sim\! t^{-2}$, is given approximately by:
\begin{equation}
F_{\rm obs}[\nu,t]  \propto [n_p]^{(1+\alpha)/2}\,\nu^{-\alpha}\propto
t^{-(1+\alpha)}\,\nu^{-\alpha}.
\label{esync}
\end{equation}
In cases for which $\nu_b(0)$ is initially well above the UVOIR bands,
 $\alpha\!\approx\! 0.5$, and the initial UVOIR  behaviour is
$F_{\rm obs}[\nu,t] \!\propto\! t^{-1.5}\, \nu^{-0.5},$ 
while the X-ray AG, for which 
$\nu_b \!\ll\! \nu$ and $\alpha\!\simeq\! 1.1$, behaves like
$ F_{\rm obs}[\nu,t] \!\propto\! t^{-2.1}\, \nu^{-1.1}$ (Dado, Dar \& De
R\'ujula~2003b). 
Very early optical light curves, declining with the expected power-law,
$\sim\! t^{-1.5}$, have been observed, e.g.,
in GRB 990123 (Akerloff et al.~1999); GRB 021211 (Li et al.~2003) 
and GRB 061126 (Perley et al.~2007). The steeper $\sim\! t^{-2.1}$
decline of the X-ray synchrotron emission is hidden under the
dominant early-time ICS light-curve, whose predicted early 
decline, $\propto\! t^{-2}$, is very similar.

Finally, in the CB model, the extinction in the host galaxy may
be time dependent.
In a day of the highly Doppler-shortened observer's time, CBs typically 
move to kiloparsec distances from their birthplace, a region wherein 
the extinction should have drastically 
diminished. A strong variation in extinction as a function of time was 
reported, e.g., by Watson et al.~(2007). Since the
extinction is chromatic and changes with the changing line of 
sight to the moving CB in the host galaxy, and since the extinction in the 
host is not well determined and usually not corrected for, it can produce 
an observed chromatic AG, even if the emitted AG had an achromatic 
light-curve.

\section{Case Studies}
\label{Case} 

To date, SWIFT has detected nearly 250 long GRBs, localized most of 
them through their $\gamma$, X-ray and UVO emissions and followed them 
until they faded into the background.  Beside the SWIFT observations, 
there have been many prompt optical measurements of SWIFT GRBs by 
an increasing number of ground-based robotic telescopes, and follow-up 
measurements by other ground-based optical telescopes, including some of 
the largest ones. Incapable of discussing all SWIFT GRBs, we discuss a 
representative set with X-ray and optical light curves which are well 
sampled and measured for relatively long times, and are tabulated in 
publications, or were made available to us following direct requests to 
authors\footnote{Recently, Evans et al.~(2007) have created an excellent 
site, with a repository of SWIFT XRT lightcurves for all SWIFT GRBs: {\it 
http://www.SWIFT.ac.uk/xrt\_curves/}}. The sample includes the brightest 
SWIFT GRBs, the one with the longest measured X-ray emission, a couple 
with canonical X-ray light curves (with and without X-ray flares) and some 
of the ones considered most puzzling from the point of view of fireball 
models.

We fit the X-ray light curves with use of Eq.~(\ref{GRBXlc}) for the ICS 
contribution and Eq.~(\ref{Fnu}) for the synchrotron emission. The 
a-priori unknown parameters are the environmental ones along the CBs' 
trajectory (the distributions of the glory's light and the ISM density),
the number of CBs, their ejection time, 
baryon number, Lorentz factor and viewing angle.
To demonstrate that the CB model correctly describes all of
the observed features of the SWIFT X-ray observations, 
 it suffices to include in the fits only the main or the latest few
observed pulses or flares. This is because the last exponential
factor in Eq.~(\ref{GRBXlc}) suppresses very fast the 
relative contribution of the earlier pulses by the time the
data sample the later pulses of flares.
It also suffices to fix the glory's light and ISM density
distribution to be the same along the trajectories of all CBs in a given GRB.  
The pulse shapes are assumed to be universal: given by Eq.~(\ref{GRBXlc}). 
For the synchroton contribution, it suffices to consider a common
emission angle $\theta$ and an average initial Lorentz factor 
$\gamma_0$ for the ensemble of
CBs, taken to be the same
as for the last two flares (or the last one, if it is unique). 

The ISM density along 
the CBs' trajectories is taken to be constant. A `windy' contribution 
($\propto\! 1/r^2$) near the ejection site is only relevant in some
optical synchrotron-dominated AGs for which very early data are
available, as discussed in the previous section.
The UVO and X-ray light 
curves are calculated with the same parameters as the X-ray light curves. 
The optical extinction and the X-ray absorption in the host galaxy 
and ours are
those required by the observations. 
The spectral index of the Fermi-accelerated electrons in 
the CBs was kept fixed $p=2.2$, i.e., $\alpha=0.5$ below the bend 
frequency and $\alpha=1.1$ above it in Eq.~(\ref{fsyn}).
The parameters used in the
CB model description of the ICS flares and the synchrotron afterglow  
of all the GRBs and XRFs to be discussed anon
are listed in Tables \ref{t1} and \ref{t2}. 

\noindent
{\bf GRB 060526.} {\it Observations:}

This GRB
was detected by SWIFT's BAT at 16:28:30 UT on May 26, 
2006 (Campana et al.~2006a). The XRT began observing the field 73 s
after the BAT trigger. The burst started with a  
$\gamma$-ray peak
lasting 18 s. The GRB was thereafter quiet for about 200 s, 
and then emitted two additional pulses which lasted about 50 s, and 
were also coincident with strong X-ray flares.
The XRT followed the X-ray emission for 6 days until it faded into 
the background. The entire XRT light curve is shown in Fig.~\ref{f3}a. It 
has the `canonical behaviour' of many  SWIFT GRBs,  
and two clear bumps.

The Watcher 40cm robotic telescope, in South Africa, caught the 
burst 36.2 s after the SWIFT trigger. It saw the AG at a very 
bright 15th magnitude. The UVOT on SWIFT detected its optical AG 81 s
after the BAT trigger. The burst was followed up with UVOT and 
ground-based telescopes by several groups. Spectra obtained with the 
Magellan/Clay telescope indicated a redshift of $z\!=\!3.21$ (Berger \& Gladders 
2006). Its R-band light curve   with the 
MDM and PROMPT telescopes at Cerro Tollolo, and with other telescopes, 
is shown in Fig.~\ref{f3}b (Dai et al.~2007 and references therein).
It can be seen in
Fig.~\ref{f3}a,d that, from the onset of the plateau, the X-ray light curve and 
the well sampled R-band light curve show, ignoring possible mini-flares,
an achromatic behaviour.

\noindent
{\bf GRB 060526.} {\it Interpretation:}

The entire XRT light curve and its CB-model's fit are shown in 
Fig.~\ref{f3}a. Three pulses were used in the fits. The ICS density was 
taken to be constant. An enlarged view of the early 
light curve is shown in Fig.~\ref{f3}b, a domain in which ICS dominates. A 
zoom-in on the two major X-ray flares at the end of the GRB is shown in 
Fig.~\ref{f3}c. The decay of the prompt emission is dominated by the decay 
of the third pulse. 
The early ICS flares, the decay of the prompt emission 
and the subsequent synchrotron-dominated plateau and gradually bending 
light curve into a power-law decay are all well reproduced by the CB model. In 
Fig.~\ref{f3}d we show the theoretical R-band light curve obtained with the 
parameters which were fitted to the X-ray light curve. Since the bending 
frequency during the steepening phase is below the R band, the temporal 
decay of the R-band light curve practically coincides with that of the 
X-ray light curve. Both late time light curves are bumpy, which may be 
caused by miniflares and/or density inhomogeneities, which we have not 
tried to fit.

\noindent
{\bf GRB 050315.} {\it Observations:}

This GRB was detected and located by BAT at 20:59:42 UT on March 15, 
2005. The BAT light curve comprises two major overlapping peaks separated 
by about 22 seconds. Absorption features in the spectrum of its  optical 
afterglow  obtained  with the Magellan telescopes indicated that its 
redshift is $z\geq 1.949$ (Kelson \&  Berger 2006). 
The XRT began observations 80 s after the 
BAT trigger and continued them for 10 days, providing one 
of the best sampled X-ray AGs (Vaughan et al.~2006). 
The extrapolation of the BAT light curve into the XRT band pass 
showed the X-ray data to be consistent with the tail end of the decaying 
prompt emission. 
The combined light curve showed the canonical behaviour: 
the rapid decline ends $\sim\!300$ s 
after trigger, the plateau lasted for about 
$10^4$ seconds, before it gradually bent into a power-law 
decay. 

\noindent
{\bf GRB 050315.} {\it Interpretation:}

The complete X-ray light curve of GRB 050315 is compared with the CB-model 
prediction in Fig.~\ref{f4}a. An enlarged view of the X-ray emission during 
the GRB and the fast decline at its end are shown in Fig.~\ref{f4}b. Two 
pulses are used in the fit.  Notice 
how snugly the model reproduces the data: the exponentially decaying 
contributions of the two pulses describe the changing slope of the 
fast-decaying phase.  The early ICS flares, the decay of the prompt 
emission and the subsequent synchrotron-dominated plateau and gradually 
bending light curve into a power-law decay are well reproduced by the 
model.

  \noindent
{\bf GRB 050319.} {\it Observations:}

This GRB  was detected and located by BAT aboard SWIFT at 09:31:18.44
UT March 19, 2005. A reanalysis of the BAT data showed that its onset was
$\sim\!135$ s before the trigger time reported by Krimm et al.~(2005).
The XRT began its observations  90 s
after the BAT trigger, continuing them for 28 days
(Cusumano et al.~2006a).
The  X-ray light curve had the canonical behaviour: 
an early fast decline which extrapolated well to the low-energy tail of 
the last prompt pulse at around 137 s after the onset of the GRB.
After $\sim\!400$ s, the fast decline was overtaken by a plateau
which gradually bent into a power-law decline after $\sim\!10^4$ s.
SWIFT's UVOT was able to observe the UVO emission 140 s 
 after its detection by the BAT. It was also observed by 
ground-based robotic telescopes:
the Rapid Telescopes for Optical Response 
 system RAPTOR (Wozniak et al.~2005),
and ROTSE III (Quimby et al.~2006) just 
27.1 s after the SWIFT  trigger.  
The optical AG was followed with a number of ground based 
telescopes (Huang et al.~2006 and references therein). An absorption 
redshift, $z\!=\!3.24$, was reported from observations with the Nordic Optical 
Telescope  (Jakobsson et al.~2006).

\noindent
{\bf GRB 050319.} {\it Interpretation:}

The CB model fit to the complete X-ray light curve and an enlarged view of 
the fit at early time are shown in Figs.~\ref{f4}c,d, respectively. Two 
pulses are used in the early ICS phase. 
The early ICS flares, the decay of the prompt emission 
and the subsequent synchrotron-dominated plateau and gradually bending 
light curve into a power-law decay are well reproduced by the model.

\noindent
{\bf GRB 060729.} {\it X-ray observations:}
 
This interesting GRB 
was detected and located by SWIFT at 19:12:29 on July 29, 2006 
(Grupe et al.~2006). It is the brightest SWIFT GRB in X-rays after GRB 061121 
and it has the longest follow-up observations in X-rays: 
more than 125 days after burst (Grupe et al.~2007). The X-ray 
observations were triggered by the detection of a GRB precursor by the 
BAT, which dropped into the background level
within 6 s. Two other major overlapping peaks 
were detected 70 s after trigger and a fourth one around 120 s. The 
end of the fourth peak was seen also by the XRT at the beginning of its 
observations. The XRT detected another flare around 180 s after which the 
light curve decayed rapidly by three orders of magnitude before it was 
overtaken by the plateau at $530 \pm 25$ s, which lasted
for $\sim\!1/2$ day before it bent into a power-law 
decline. The complete light curve obtained from the 
observations with the BAT (extrapolated to the XRT band), the XRT and 
XMM-Newton is shown in Fig.~\ref{f5}a; the early times are shown
in Fig.~\ref{f5}b. 

\noindent
{\bf GRB 060729.} {\it Interpretation:}

The CB model fit to the complete X-ray light curve of GRB 060729 with just 
the two most prominent X-ray peaks is shown in Fig.~\ref{f5}a.  
In Fig.~\ref{f5}b all five flares are modeled.
The overall good agreement in Fig.~\ref{f5}a extends over some
eight orders of magnitudes in time and in flux. 
The spectral evolution of the X-ray emission is also in good agreement 
with the CB-model predictions. In the XRT 0.3-10 keV band the spectrum of 
the early time flares and their spectral evolution are well described by 
the broken power-law obtained by ICS of a thin bremsstrahlung spectrum, 
Eq.~(\ref{GRBXlc}). In particular the exponential factor in Eq.~(\ref{GRBXlc}) 
describes well the rapid softening of the spectrum as a function of time 
during the fast decaying phase of the flare. But, as soon as the X-ray 
afterglow is taken over by the synchrotron emission around 325 s, the  
predicted spectral index is $\beta_X\!=\!1.1$, in excellent agreement with 
the observations of the XRT and of XMM-Newton (Grupe et al.~2007).

\noindent
{\bf GRB 060729.} {\it Optical observations and a sketch of their
interpretation:}

The ROTSE-IIIa telescope in Australia took a first 5 s image of this GRB,
starting about a minute after the burst, which showed no afterglow 
down to magnitude 16.6. Some 23 s later, an AG of magnitude 
15.7 was clearly detected. The AG brightened over several minutes, 
and faded very slowly. Such a behaviour is expected from the combined 
ICS and synchrotron emission in the UVOIR band 
during and shortly after the prompt GRB. The UVOT followed the UVO 
emission  from 739 s after trigger until 20 days after burst. 
The VLT in Chile obtained spectra, and determined a relatively low 
redshift of $z\!=\!0.54$ for this burst (Theone et al.~2006). 
The light curves of its UVO AG show a striking similarity to 
the X-ray light curve (Grupe et al.~2007) as predicted by the CB model for 
the optical AG when the bending frequency is below the UVO band, and the 
extinction along the lines of sight to the hyperluminal CBs stays constant.
All in all, the well sampled  observations
of the XUVO light curves of GRB 060729 agree well with the 
expectations of the CB model.

\noindent
{\bf GRB 061121.} {\it Observations }

On rare occasions, such as when SWIFT triggers on a precursor to the 
main burst, the prompt emission and the afterglow can be observed 
at X-ray and UVO wavelengths with SWIFT's XRT and UVOT, and 
with ground based robotic telescopes. This GRB was such a case. 
It is the brightest GRB in X-rays 
observed to date by the XRT (Page et al.~2007). The BAT triggered on its 
precursor at 15:22:29 UT on November 21, 2006. Its main burst began 60 s 
after, and consisted of three 
overlapping peaks of increasing brightness, 
some   63, 69 and 74 s after trigger, as one can see in the insert
to Fig.~\ref{f2}d.
The $\gamma$-ray emission decayed fast after 75 s and became 
undetectable by the BAT beyond 140 s. The burst was also detected by 
Konus-Wind. 

The spectrum and spectral evolution of this GRB were well fit with a broken 
power-law. Its `peak' energy appeared to increase during the rise of
each flare and decreased as their flux decayed. But its isolated main strong flare 
at $\sim\! 74$ s, as in many cases studied before, had a maximum $E_p$ 
at its beginning, which decayed monotonically thereafter. 
After the bright burst, the X-ray emission ---measured by the XRT and 
later also with XMM-Newton--- began to follow the `canonical' decay.
Superimposed on the initial rapid decay from the major flare, 
are two smaller flares around 90 s and 125 s. The rapid decay is taken over 
by the plateau around 200 s and gradually breaks into a power-law 
decline, with an asymptotic power-law index -1.53 (+0.09 / -0.04).

\noindent
{\bf GRB 061121.} {\it Interpretation}

The continuously decreasing $E_p$ during the main flare
is as predicted by the CB model (DD2004). The alleged increase of 
$E_p$ during the rise time of the smaller flares is probably an 
artifact of overlapping peaks,
wherein the decay of a previous flare is taken over by 
a new flare only near its peak time. A comparison between the observed
complete X-ray light curve (Page et al.~2007) and the CB model's fit
is shown in Fig.~\ref{f5}c. An enlarged view of the early time behaviour 
is shown in Fig.~\ref{f5}d. The general trend before the onset
of the synchrotron plateau is dominated by the main X-ray peak.
The two smaller overlapping preceding peaks seen in the
$\gamma$-ray light curve in Fig.~\ref{f2}d have been included
in the fit, and so have the two late X-ray flares not intense enough
to be seen in $\gamma$-rays. The
model reproduces very well the observed  light curve  over seven
 orders of magnitude in intensity and five orders of magnitude in time.  
The late temporal-decay index is
in good agreement with the universal power-law index, $-1.6$, predicted by 
the CB model for constant ISM density. 

The CB model also correctly predicts the spectrum and spectral evolution 
of the X-ray emission during the rapid decline and subsequent 
AG phase. The observed strong softening of the 
spectrum during the rapid-decline
is in full agreement with the spectral evolution predicted 
by the CB model, Eq.~(\ref{GRBXlc}). When the plateau phase takes over, 
the spectral power-law index changes to $-1.07\pm 0.06$, and remains in 
agreement with the  prediction for the synchrotron AG, 
$\beta_X=-1.1$. The slight hardening of the spectrum at late time
to $\beta_X=0.87 \pm 0.08$ we have not predicted. It may
be due to the contribution from ICS of the microwave background 
photons by cosmic ray electrons accelerated by the CBs (Dado
and Dar, 2006; 
Dado, Dar \& De R\'ujula, to be published).
 
\noindent
{\bf GRB 060206.} {\it Observations}
 
This GRB triggered the SWIFT's BAT on February 6, 2006 at 04:46:53
UT (Morris et al.~2006). Its $\gamma$-ray emission lasted only 
6 s. The XRT started its observations 80 s after the BAT trigger.
Despite its initial poor time sampling,  it detected  
an X-ray decline after 1/2 hour and a strong rebrightening after one hour,
after which its follow up was nearly continuous for some 20 days.
The bright optical AG of GRB was detected by SWIFT at $V=16.7$ 
about a minute after the burst.   
RAPTOR   started  observations 
48.1 minutes after the BAT trigger and reported that after 
an initial fading, the AG rebrightened 1h after the burst 
by $\sim\! 1$  magnitude within a couple of minutes 
(Wozniak et al.~2006). Many observatories followed the bright optical AG
(Stanek et al.~2007 and references therein), and Fynbo et 
al.~(2006a) carried out spectral observations to determine its 
large redshift, $z=4.05$, later confirmed by other groups.
The RAPTOR data clearly shows that the rebrightening was due to a couple 
of flares (Wozniak et al.~2006). Similar ``anomalous'' rebrightening  of the
optical afterglow was seen in some other bursts (Stanek et al.~2007) .

\noindent
{\bf GRB 060206.} {\it Interpretation}

In Fig.~\ref{f6}a,b we compare the observations of the X-ray
and R-band light curves with the CB-model fits. 
Superimposed on the plateau 
phase are two strong flares beginning around 1 h after trigger. 
In Fig.~\ref{f6}c we compare the observed
light curve of these two flares in the R-band 
and their CB model description via Eq.~(\ref{GRBXlc}). The figures show
that the agreement is very good and that there is 
nothing `anomalous' in the X-ray and optical data of GRB060206.
Instead,
their prominent structures are well described and precisely
related by their CB-model's understanding in terms of ICS.

\noindent
{\bf XRF 050406.} {\it Observations}

This XRF was triggered by the SWIFT's BAT 
on April 6, 2005 at 15:58:48 UT (Parsons et al.~2005). It was a 
$\sim\! 5$ s burst with a soft power-law spectrum and no significant emission 
above 50 keV, implying its classification as an XRF (Heise et al.~2001).
It was the first burst detected by SWIFT showing an early 
time ($t\sim\! 210$ s) X-ray flare after the decay of its $\gamma$-ray 
emission, and the first XRF with a well studied early-to late X-ray light 
curve. The XRT and UVOT began an automated series of observations of its 
BAT error circle 88 s after the BAT trigger. The X-ray and UVO afterglow 
were not detected on board, but were discovered in on-ground analysis.

\noindent
{\bf XRF 050406.} {\it Interpretation}

In Fig~\ref{f7}a we present a comparison between the X-ray
light curve and a CB model fit with a single X-ray flare with 
$t_p \!=\! 210$ s. In Fig~\ref{f7}b we add a second flare at $t_p \!=\! 
1000$ s. The agreement between theory and observations is good,
but not as impressive as in other cases, given
the relatively poor observational sampling.

\noindent
{\bf XRF060218/SN2006aj.} {\it Observations}

This XRF/SN pair provides one of the best testing grounds of theories,
given its proximity, which resulted in very good sampling and statistics.
The XFR was detected with the SWIFT's BAT on 2006 February 18, at 
03:34:30 UT (Cusumano et al.~2006b). The XRT and UVOT detection  
(Marshall et al.~2006; Campana et al.~2006b) led to its 
precise localization, determination 
of its redshift, $z\!=\!0.033$, (Mirabal \& Halpern, 2006) and the 
discovery of its association with a supernova, SN2006aj (Nousek 
et al.~2006).  The BAT data lasted only 300 s, beginning 159 s after 
trigger, with most of the emission below 50 keV (Campana et al.~2006b; 
Liang et al.~2006). The combined BAT and XRT data showed that the prompt 
$\gamma$-ray and X-ray emission lasted more than 2000 s, with a total 
isotropic equivalent $\gamma$-ray energy, $E_{\rm iso}\sim 0.8\times 10^{49}$ 
erg and a spectral peak energy, $E_p$, which strongly evolved with time 
from 54 keV at the beginning down to $\sim$5 keV at later times.

\noindent
{\bf XRF060218/SN2006aj.} {\it Interpretation}

The spectral energy distribution as measured by 
the BAT and XRT was modeled (e.g.~Campana et al.~2006b; Liang et al.~2006)
by a sum of a black body emission with a 
time-declining temperature and a cut-off power-law. 
This alleged black body emission was 
interpreted as the result of the core-collapse shock breaking out from the 
stellar envelope (Colgate 1968), and of the stellar wind of the progenitor 
star of SN2006aj (Campana et al.~2006b; Blustin 2007; Waxman, Meszaros \& 
Campana, et al.~2007). From this interpretation, a delay of $\leq 4$ ks 
between the SN and the GRB start was concluded.
Amati et al.~(2006) claimed that XRF060218 complies with the so-called 
`Amati relation' for GRBs and XRFs and indicates that XRF 060218 was not a 
GRB viewed far off axis. 

The association of XRF060218 with SN2006aj,  a Type Ic supernova, akin to
the ones of the GRB980425/SN1998bw  (Galama et al.~1998),
GRB030329/SN2003dh (Stanek et al.~2003; Hjorth et al.~2003) and
GRB 031203/SN2003lw (Malesani et al.~2006) pairs, supports the 
CB-model understanding of XRFs as ordinary GRBs viewed at 
much larger angles (DD2004, Dado et al.~2004).
In this model, the isotropic equivalent $\gamma$-ray energy
emission of a typical CB is $\approx 0.8 \times 10^{44}\,[\delta_0]^3$ erg
(DD2004). 
Thus, the reported $E_{\rm iso}\approx (6.2\pm 0.3)\times 10^{49}$ erg 
implies that the CB which generated the single peak of
XRF 060218 had $\delta_0\!\sim\! 92$. It then follows from Eq.~(\ref{ICST})
that its early time $E_p=54$ keV implies (for the typical
$\epsilon_g\!\sim\! 1$ eV and
$1\!+\! \cos\theta_i\!\sim\! 0.5 $) a Lorentz factor $\gamma_0\!\sim\! 910$ and a 
viewing angle $\theta\!\sim\! 4.8\times 10^{-3}$. 
Using these parameters and a fit value for the ISM density
we can predict via Eqs.~(\ref{decel},\ref{Fnu}) the shape
of the plateau and the subsequent decline of the
X-ray light curve in the 0.3-10 keV band.   
The values of $\gamma_0$
and $\theta$ deduced, as above, from the prompt value of $E_p$
were used in calculating   the initial ICS-dominated X-ray peak.
The normalization, peak time and width parameters of the initial ICS-dominated
X-ray peak  as listed in Table~\ref{f2} are compatible with the BAT light curve 
and Eq.~(\ref{widthrelation}).

In Fig.~\ref{f7}c we compare the theory to the light curve 
inferred from the BAT and XRT observations. With use of
the CB model pulse shape, we can infer from either the BAT background 
count-rate or the XRT light curve that the GRB pulse started some
$10^3$ s before the  first BAT image of the burst. 
The agreement between the observed and the predicted light curves is very 
good at all times. In particular, the exponential factor in 
Eq.~(\ref{GRBXlc}) describes very well the fast-decaying phase of the XRF,
and the fast softening of its spectrum
during this phase. As soon as the synchrotron emission dominates 
over ICS (around $\sim 9000$ s) the spectrum 
becomes a harder power-law with an observed index
$\beta_X\sim -1.1$, the prediction of the CB model for the 
unabsorbed synchrotron spectral energy density in the  X-ray band 
(DDD2002a).

The light curves of  XRF060218/SN2006aj obtained with different filters by
SWIFT's UVOT  are particularly interesting. Not only they  provide
evidence that the XRF was produced in the explosion of SN2006aj, but 
they confirm the CB-model interpretation of the broad band emission at  
all times. Prior to the dominance of 
the associated supernova's radiation, the UVO light curves show wide peaks whose 
peak-time shifts  from $t_{peak}\approx 30000$ s at 
$\lambda\sim 188$ nm to $t_{peak}\approx 50000$ s at $\lambda\sim 544$ nm, see Fig.~\ref{f1}.
Moreover, their peak-energy flux decreases with energy. 
In the CB model these are the predicted properties
 of a single peak generated
by a single CB as it Compton up-scatters light of various initial
angles of incidence and energy
in the spectrum of the glory's light. We proceed to
prove this point for the observed peak-energy flux and
peak-times in the different UVOT filters.

The spectrum of Eq.~(\ref{GRBspec}) 
with $\alpha\!\sim\! 1$,  and $E\!\ll\! T$,  is approximately
constant: independent of frequency. Using the approximation
$\Delta\nu\!\approx\! \, c\,\Delta\lambda/\lambda^2$, we can predict the
integrated fluence in the various UVOT bands from that in the X-ray
energy interval 0.3-10 keV at its $t_{peak}\sim 870$ s. We correct the results 
for extinction in the Milky Way, $E(V\!-\!B)\!=\!0.127$; and in the host-galaxy, 
$E(V\!-\!B)\!=\!0.044$ (Guenther et al.~2006). In Table~\ref{t3} we compare
the predictions with the observations. The error in the predictions
is the combined error of the extinction correction and of the observed
peak flux in the XRT band. The results are excellent.
The flux ratios would have been completely different for a 
thermal black body. Thus, we see no evidence 
in the observations for a significant contribution from an alleged thermal 
black body source (e.g., Campana et al.~2006b, Liang et al.~2006).

The $E\!\times\! t^2$ law of Eq.~(\ref{law}) implies that the peak time
of the energy flux of a pulse observed at various energies
scales approximately as $E^{-1/2}$.
This is tested in Table~\ref{t4} and in 
Fig.~\ref{f8}b, where we compare the predicted 
and observed values of $t_{peak}$.
In the comparison we used $t_i\!\approx \!-\!1080$ s,
obtained from the CB model description
of the XRT light curve, and consistent with the back-extrapolation 
of the rising part of the 
BAT light curve to the earlier background level.
The agreement between data and observations is excellent.

In Fig.~\ref{f8}a we reproduce our earlier study of the same 
$E\!\times\! t^2$ law, tested in DD2004 for the average full-width at 
half-maximum (FWHM) of 
the peaks of ensembles of GRBs of undetermined redshift
in the four BATSE energy channels (Fenimore et al.~1995; 
Norris et al.~1996). This earlier test 
encompassed a one-order-of-magnitude energy interval, 
its extension to two extra orders of magnitude in
Fig~\ref{f8}b is another very satisfactory test of the CB-model's ICS
mechanism for the origin of the `prompt' GRB radiation.
For an XRF such as 060218, the optical peaks are observable
and not so `prompt' at all. This is a straightforward consequence of
the CB-model's GRB/XRF unification, and of the relatively large observer's angle of XRFs. A more detailed analysis of XRF060218/SN2008aj 
will be published elsewhere.

We conclude that this XRF/SN pair is in full agreement with the predictions
of the CB model. The rich structure of its UVO AG is as expected.
Its X-ray light curve has the canonical shape seen in ordinary GRBs.
Both of these results are explicitly dependent on the fact that XRFs
are GRBs viewed at a larger angle. The data do not 
require a black body component produced by shock break out through the 
stellar envelope, or by any other mechanism. The start time of the X-ray 
emission does not constrain the exact time of the core's collapse before the 
launch of the CBs, nor does it constraint possible ejections of other CBs 
farther off axis, prior\footnote{Interestingly, 
SWIFT detected gamma rays from the same 
direction over a month earlier on January 17, 2006 - 
http://grb.sonoma.edu/ } to the trigger-time of XRF060218.

\section{Conclusions and outlook}
\label{outlook}

The data on GRBs gathered after the launch of SWIFT, as interpreted
in the CB model and as we have discussed
here and in recent papers (Dado et al.~2006; 2007a; 2007c) 
has taught us four things:
\begin{itemize}
\item{}The relatively narrow pulses of the $\gamma$-ray signal, the
somewhat wider prompt flares of the X-rays, and the much wider humps
sometimes seen at UVOIR frequencies, have a common origin. They are
generated by inverse Compton scattering.
\item{} The historical distinction between prompt and afterglow phases
is obsolete. It is replaced
by a physical distinction: the relative dominance of the Compton or
synchrotron mechanisms at different, frequency-dependent times.
\item{} The two mechanisms quoted above: Compton and
synchrotron, suffice to provide a very simple and accurate description
of XRFs and long-duration GRBs at all frequencies and times.
Simple as they are, the mechanisms generate the rich structure
of the light curves at all frequencies, and their chromatic or
achromatic `breaks'.
\item{} To date, as the quality of the data improves, so does the 
 quality of its agreement with the CB-model's predictions.
\end{itemize}

We re-emphasize that the 
results presented in this paper are  based on direct applications
of our previously published explicit predictions.
Our master formulae, Eq.~(\ref{GRBXlc}) for ICS and 
Eqs.~(\ref{decel}, \ref{Fnu})
for the synchrotron component describe all the data very
well. But, could they just be very lucky guesses? The general 
properties of the data are predictions. But,
when fitting cases with many flares, are we not `over-parametrizing' 
the results? Finally, the $E\!\times\!t^2$ law plays an important role.
Could it be derived\footnote{Such a correlation has
been discussed in FB models. Wu and Fenimore (2000)
studied it as a consequence of synchrotron cooling,
concluding that `it cannot simultaneously explain
the large flux without a large magnetic field
and the [correlation] without a smaller one'. Recently,
Liang et al.~(2007) studied it in the case of XRF060218,
concluding that `after subtracting the thermal component',
the correlation is $t\!\approx\! E^{-0.25}$.}
in a different theory? 
To address questions such as these, it may be
 useful to place the results of this 
paper in the context of our previous work on the subject.

In  DDD2002a and DDD2003a
we demonstrated that the synchrotron mechanism --generally
dominating  the `afterglows' of long GRBs and XRFs at late times-- 
provides an excellent description of the observations, all the way from 
X-rays to radio, and with no exceptions.
The bolometric AG energy, the spectral shapes and 
 evolution, including the smooth achromatic temporal 
 `breaks' observed in the XUVOIR light 
curves, are all simple  consequences of the theory.
The `canonical' behaviour of X-ray light curves was predicted.

The precise understanding of the AG generated by the approaching
jet of CBs, allowed us to `see' the associated supernova 
(as opposed to an `echo' enhancement)
in all cases where the data were sufficient
accurate, essentially all cases with $z<1.1$
(DDD2002a). This convinced us of two facts: that core-collapse supernovae 
are the `engines' of XRFs and long GRBs, as hypothesized in
Dar and Plaga (1999), Dar and De R\'ujula (2000a)
and DDD2002a; that an
otherwise normal SN, if seen nearly on axis, 
appears to have  `peculiar' properties akin to those of SN1998bw, our 
adopted `standard candle'.
Based on these 
results, we could predict in several cases the appearance of a SN in a GRB
afterglow (Dado, Dar \& De R\'ujula 2002b,c). One spectacular case concerned 
the SN associated with GRB030329, for which we even predicted the exact
date of its spectroscopic discovery (Dado et al.~2003d). The 
GRB030329/SN2003dh
pair (Stanek et al.~2003; Hjorth et al.~2003)
changed  the `standard'  view. Nowadays
the surprise is to find a GRB such as 060614, arguably
(Schaefer 2006, Dado et al.~2006) not associated with a 
SN1998bw-like supernova (Gal-Yam et al. 2006).

Given the information extracted from our AG analyses on the typical
Lorentz factors, observation angles and cannonball masses
of pre-SWIFT GRBs,
the observed properties of pre-SN winds, and the early luminosity of
a core-collapse SN (such as SN1987A),
we  went one step forward. We studied the outcome
of inverse Compton scattering of the light of the supernova's
`glory' by the electrons enclosed in a CB (DD2004). As it turns out,
the mechanism correctly predicts all the established
properties of GRB pulses, their typical values, and the distributions
around them. These include their peak-energies, 
equivalent isotropic energy, pulse shape, 
spectrum and temporal evolution. 

A tell-tale signature of ICS
is a large polarization of the prompt radiation
(Shaviv \& Dar 1995), predicted to be smaller
in XRFs (DD2004; Dado et al.~2007), 
and negligible in the AGs of both XRFs and GRBs
(DD2004, Dado et al.~2007b). 
A large polarization was reported in the prompt
radiation of four GRBs (021206: Coburn \& Boggs~2003;
930131, 960924:  Willis et al.~2005;  
041219a: Kalemci et al.~2006; McGlynn et al.~2007).
But the data are controversial:
the errors are large and the extraction of a signal is difficult.

The correlations between GRB observables have attracted much interest.
They were predicted in Dar and De R\'ujula (2000b), further discussed and 
confronted with early data in DD2004. In Dado et al.~(2007a)
 we have studied the
correlations expected in the CB model in more detail, and successfully
confronted them with recent data (Schaefer 2006 and references 
therein). These correlations are amongst the most 
straight-forward consequences of the ICS mechanism. More recently,
correlations between prompt and AG observables, as well as
between two AG observables (the total energy in an X-ray band
and the ending time of its plateau phase) have been studied
(Nava et al.~2007).
The observations precisely agree with the CB-model's
predictions (Dado et al.~2007c). In particular, the relation between
the total isotropic energies in the prompt and AG phases was never
a problem in the CB model, which correctly predicted both of these
energies.
In Dado et al.~(2007c) we have also shown that their distributions
and correlations are as expected.

When their collimated radiation points to the observer, GRBs 
are the brightest sources in the sky.
In the context of the CB model and of the simplicity of its underlying physics, GRBs are not  persistent mysteries,
`the biggest of explosions  after the Big Bang', and a constant source of surprises, exceptions and new requirements. 
Instead, they are well-understood and can be used as cosmological tools, 
to study the history of the intergalactic medium and of star formation up to 
large redshifts, and to locate SN explosions at a very early stage.
As interpreted in the CB model, GRBs are not `standard candles',
their use in `Hubble-like' analyses would require further
elaboration. The GRB conundra have been reduced to just one:
`how does a SN manage to sprout mighty jets?'
The increasingly well-studied  ejecta of quasars and microquasars,
no doubt also fired in 
catastrophic accretion episodes on compact central objects, 
provides observational hints with which, so far, theory
and simulations cannot compete.

The CB model underlies a unified theory of high energy astrophysical phenomena. The information gathered in our study of GRBs can be
used to understand, also in very simple terms, other phenomena.
The most notable is (non-solar) Cosmic Rays. We allege  (Dar 
et al.~1992; Dar \& Plaga~1999)  
that they are simply the charged ISM particles scattered by 
CBs, in complete analogy with the ICS of light by the same CBs.
This results  in a successful
description of the spectra of all primary cosmic-ray
nuclei and electrons at all observed energies
(Dar and De R\'ujula~2006a). The CB model also predicts very
simply the spectrum of the gamma background radiation 
and explains its directional properties (Dar \& De R\'ujula~2001a; 2006b).
Other phenomena understood in simple terms include the
properties of cooling core 
clusters (Colafrancesco, Dar \& De R\'ujula~2003) and 
of intergalactic magnetic fields 
(Dar \& De R\'ujula~2005) . The model may also have a say in
`astrobiology' (Dar et al.~1998; Dar \& De R\'ujula~2001b).

Finally, if cannonballs are so pervasive, one may ponder
why they have not been directly seen. After all, particularly in 
astrophysics, {\it seeing is believing}. The answer is simple.
Cannonballs are tiny astrophysical objects: their typical mass is 
half of the mass of Mercury.
Their energy flux at all frequencies is $\propto\!\delta^3$,
large only when their
Lorentz factors are large. But then, the radiation is also
extraordinarily collimated, it can only be seen nearly on-axis.
Typically, observed SNe are too far to {\it photograph} their CBs
with sufficient resolution.

Only in two SN explosions that took place close enough, the
CBs were in practice observable. They were observed. One case
was SN1987A, located in the Large Magellanic Cloud,
whose approaching and receding CBs were
photographed by Nisenson and Papaliolios (2001). 
The other case was SN2003dh, associated with GRB030329,
and located at redshift $z=0.1685$. In the CB model interpretation,
its two approaching CBs were first `seen'
as the two-peak $\gamma$-ray light curve and the two-shoulder
AG (Dado et al.~2003d). This allowed us 
to estimate the time-varying angle of their superluminal
motion in the sky. Two sources or `components'
were indeed clearly seen in radio observations
at a certain date, coincident
with an optical AG rebrightening. We claim
that the observations agree with our expectations\footnote{The
size of a CB is small enough to expect its radio image to
scintillate, arguably more than observed (Taylor et al.~2004).
We only realized a posteriori that the ISM electrons a CB
scatters, synchrotron-radiating in the ambient magnetic field, would
significantly contribute at radio frequencies, somewhat blurring the 
CB's radio image (Dado, Dar \& De R\'ujula 2004b,
Dado \& Dar 2005).},  including 
the predicted inter-CB superluminal separation 
(Dar \& De R\'ujula 2000a, DD2004).
The observers claimed the contrary, though the statistical
evidence for the weaker `second component' is $>20\sigma$,
and it is `not expected in the standard model' (Taylor et al.~2004).
The no-doubt spectacular
radio picture of the two superluminally moving sources has, to our
knowledge, never been published. This is too bad, for {\it a picture is
worth a thousand words.}

{\bf Acknowledgment:}
We would like to thank S. Campana, G. Cusumano, D. Grupe, K.~L.~Page, 
and S.~Vaughan for making available to us the tabulated data of 
their published X-ray light curves of SWIFT GRBs.
This research was supported in part by the Asher Fund for Space Research 
at the Technion.

\newpage

\begin{deluxetable}{llcc}
\tablewidth{0pt}
\tablecaption{CB model afterglow parameters}
\tablehead{
\colhead{GRB/XRF} & \colhead{$\gamma_0$} & \colhead{$\theta\,{\rm [mrad]}$} &
\colhead{$n[10^{-2}\, {\rm cm}^{-3}]$}}
\startdata
060526 & 814  & 1.14 & 5.57 \\
050315 & 1162 & 0.83 & 0.20 \\
050319 & 721 & 0.26  & 1.42 \\
060729 & 1161 & 0.83 & 0.10 \\
060206 & 1279 & 0.81 & 1.23 \\
061121 & 1263 & 1.01 & 4.40 \\
050406 & 695 & 2.16  & 0.35 \\
060218 & 911 & 4.70  & 1.58 \\
\enddata
\label{t1}
\end{deluxetable}

\begin{deluxetable}{llcccc}
\tablewidth{0pt}
\tablecaption{Time parameters of the last two major X-flares}
\tablehead{
\colhead{GRB/XRF} & Band 
& \colhead{$t_1$} & \colhead{$\Delta t_1$}
&\colhead{$t_2$} &
\colhead{$\Delta t_2$}}

\startdata

060526  &  X & 233  s  & 16.4 s  &  272 s   & 31.6 s      \\
050315  &  X & -5.0 s  &  6.9 s  & 16.4 s   &  5.4 s      \\
050319  &  X & 107  s  &  7.4 s  &  137 s   &  6.2 s      \\
060729  &  X & 122  s  &  4.2 s  &  153 s   & 19.1 s      \\
060206  &  X & 581  s  & 43.2 s  & 4187 s   &  549 s      \\
061121  &  X & 68   s  &  3.8 s  &  112 s   &  4.9 s      \\
050406  &  X & 153  s  & 19.1 s  &  171 s   & 19.3 s      \\
060218  &  X &-1080 s   & 1977 s & ${\rm\;\; one}$&${\rm\!\!\! flare}$

\enddata
\label{t2}
\end{deluxetable}

\newpage

\begin{deluxetable}{llcccc}
\tablewidth{0pt}
\tablecaption{Peak Energy Flux (PEF) from XRF060218 in the UVOT 
filters}
\tablehead{
\colhead{Filter} & \colhead{$\lambda$}      & \colhead{E(eff)} &
\colhead{Band-} & \colhead{Observed PEF }  & \colhead{CB model PEF} \\
\colhead{} & \colhead{[nm] } &\colhead{[eV] } &\colhead{width}
&\colhead{${\rm [erg\, cm^{-2}\, s^{-1}]}$} & \colhead{${\rm [erg\, 
cm^{-2}\, s^{-1}]}$}  }
\startdata

XRT & 0.25 & $4900$& 0.3-10 & $(1.30\pm 0.07)\times 
10^{-8}$ & 
$(1.30\pm 0.07)\times 10^{-8}$  \\  
             &          &     & [keV]     & (input)  &        \\
         
UVW2  & 188 & 6.60 & 76 nm & $(2.29\pm 0.30)\times 10^{-12}$ &
$(2.26\pm 0.23)\times 10^{-12}$ \\  
UVM1  & 217 & 5.71 & 51 nm & $(1.20 \pm 0.10)\times 10^{-12}$ & 
$(1.14\pm0.12)\times 10^{-12}$ \\  
UVW1  & 251 & 4.94 & 70 nm & $(1.17 \pm 0.12)\times 10^{-12}$ & 
$(1.17\pm 0.12) \times 10^{-12}$ \\  
U     & 345 & 3.59 & 88 nm & $(8.80\pm 0.85)\times 10^{-13}$ & 
$(7.96\pm 0.80)  \times 10^{-13}$ \\  
B     & 439 & 2.82 & 98 nm & $(5.89 \pm 0.59)\times 10^{-13}$ & 
$(5.34\pm 0.53) \times 10^{-13}$ \\  
V     & 544 & 2.28 & 75 nm & $(2.56\pm 0.26) \times 10^{-13}$ & 
$(2.63\pm 0.26)\times 10^{-13}$ \\

\enddata
\label{t3}
\end{deluxetable}

\begin{deluxetable}{llcc}
\tablewidth{0pt}
\tablecaption{Peak time of the energy flux from XRF060218 at different
energies}

\tablehead{
\colhead{Band}   & \colhead{E(eff) [eV]} &
\colhead{Observed $t_{peak}$ [s]} & \colhead{CB Model $t_{peak}$ [s]}
}
\startdata
15-150 keV   & 4900    &  $ 425\pm 25 $    & $425\pm 25$ \\
             & (input)         &     (input)          &                   \\
0.3-10 keV   & 2100   & $985 \pm 25  $   & $ 980\pm 50$ \\
UVW2  & 6.60  & $31,300 \pm 1,800 $  & $ 27,600 \pm 1,700$ \\
UVM1  & 5.71  & $41,000 \pm 8,000 $  & $ 29,730 \pm 1,820$ \\
UVW1  & 4.94  & $41,984 \pm 8,000 $  & $ 32,000 \pm 1,970$ \\
U     & 3.59 & $42,864 \pm 3,000 $  & $ 39,360 \pm 2,500$ \\
B     & 2.82 & $39,600 \pm 2,500 $  & $ 42,560 \pm 2,600$ \\
V     & 2.28 & $47,776 \pm 3,500 $  & $ 47,400 \pm 2,900$

\enddata

\label{t4}
\end{deluxetable}

\newpage 
\begin{figure}[]
\vspace{-.7cm}
 \hskip -2 cm 
\epsfig{file=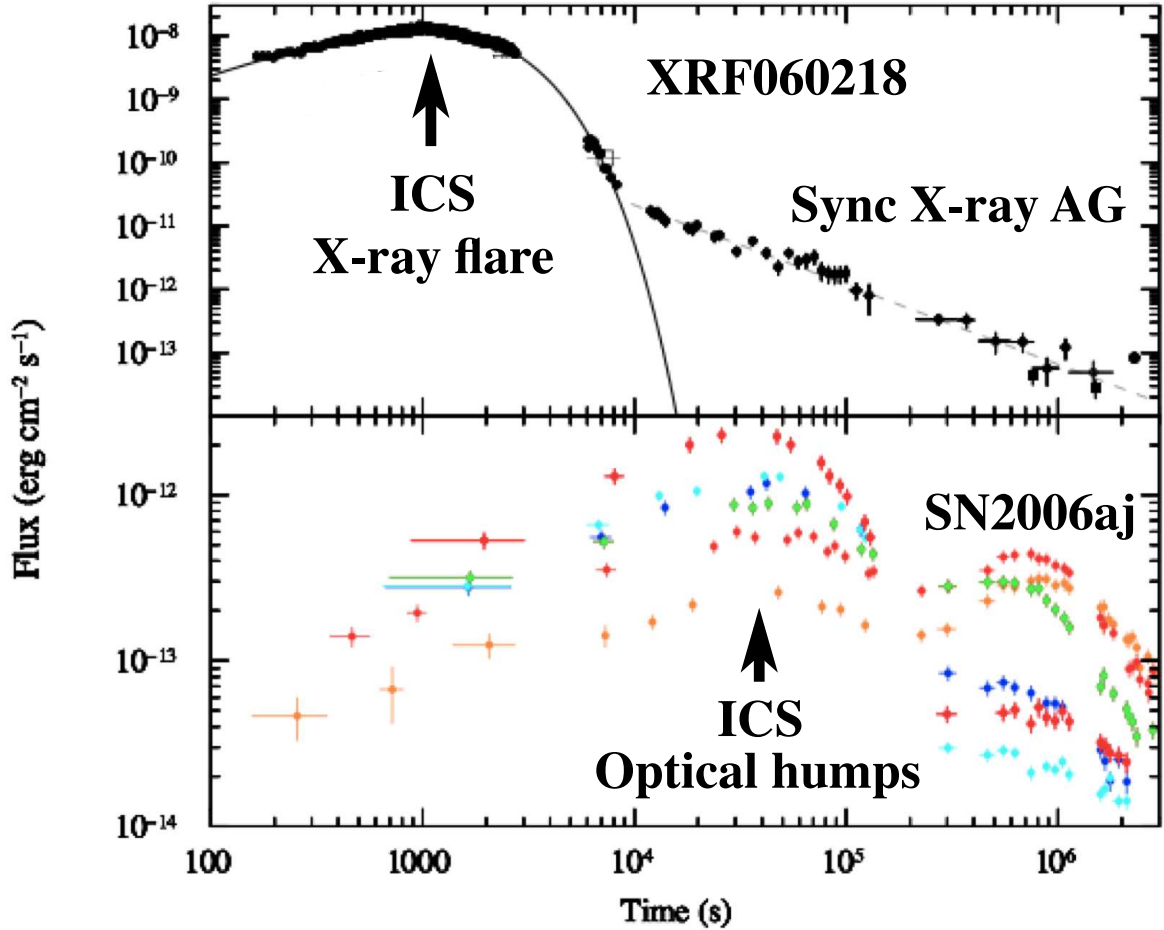,width=22cm} 
\vspace{-1cm}
\caption{Data on XRF 060218/SN2006aj.
{\bf Top}: The 0.3-10 keV SWIFT-XRT light curve.
  The black line 
is a sum of a cut-off power-law flux and a black body flux with 
an amplitude and a temperature which evolve with time
and are adjusted to fit the data
 (Campana et al.~2006b). The 
dashed line is their best fit power-law decline beyond 10000 s. The thick 
arrow indicates the  peak-flux time. 
{\bf Bottom}: UVO light curves (Campana 
et al.~2006b) with various filters: red - V 
(centered at 544 nm); green - B (439 nm); blue - U (345 nm), light blue - 
UVW1 (251 nm); magenta - UVM1 (217 nm) and yellow - UVW2 (188 nm). 
The UVO peak time (roughly indicated by an arrow) increases 
with wave length from 
nearly 30000 s for $\lambda\!\sim\! 188$ n to nearly 50000 s for 
$\lambda\sim 544$ nm. After 2 days the light curves are overtaken by the 
light from SN 2006aj, which peaks around 10 days after the BAT trigger.
}
\label{f1}
\end{figure}

\newpage
\begin{figure*}
\begin{tabular}{cc}
  \hskip -1.5 cm
  \epsfig{figure=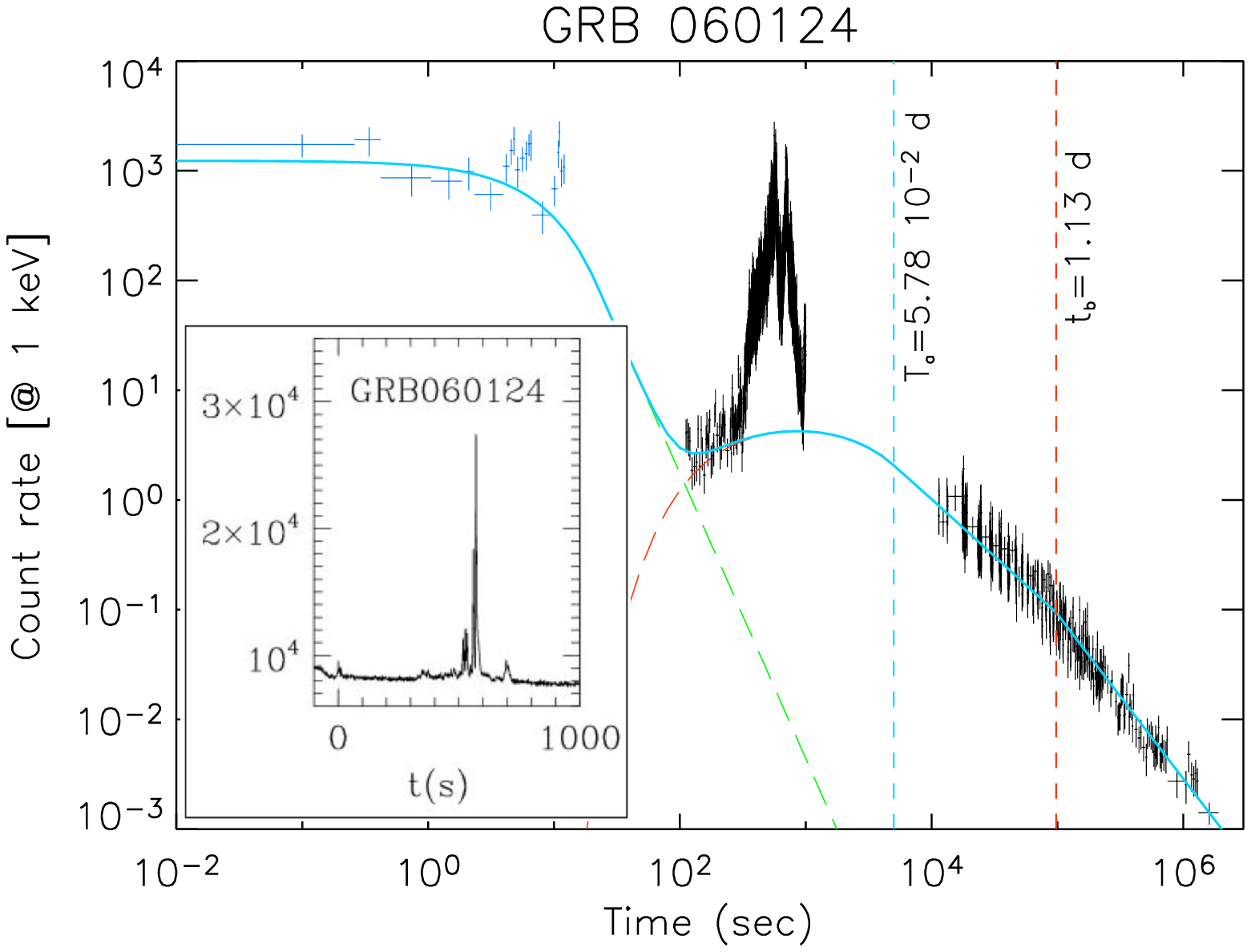,width=10.0cm}
   \hskip -1.5 cm
  \epsfig{figure=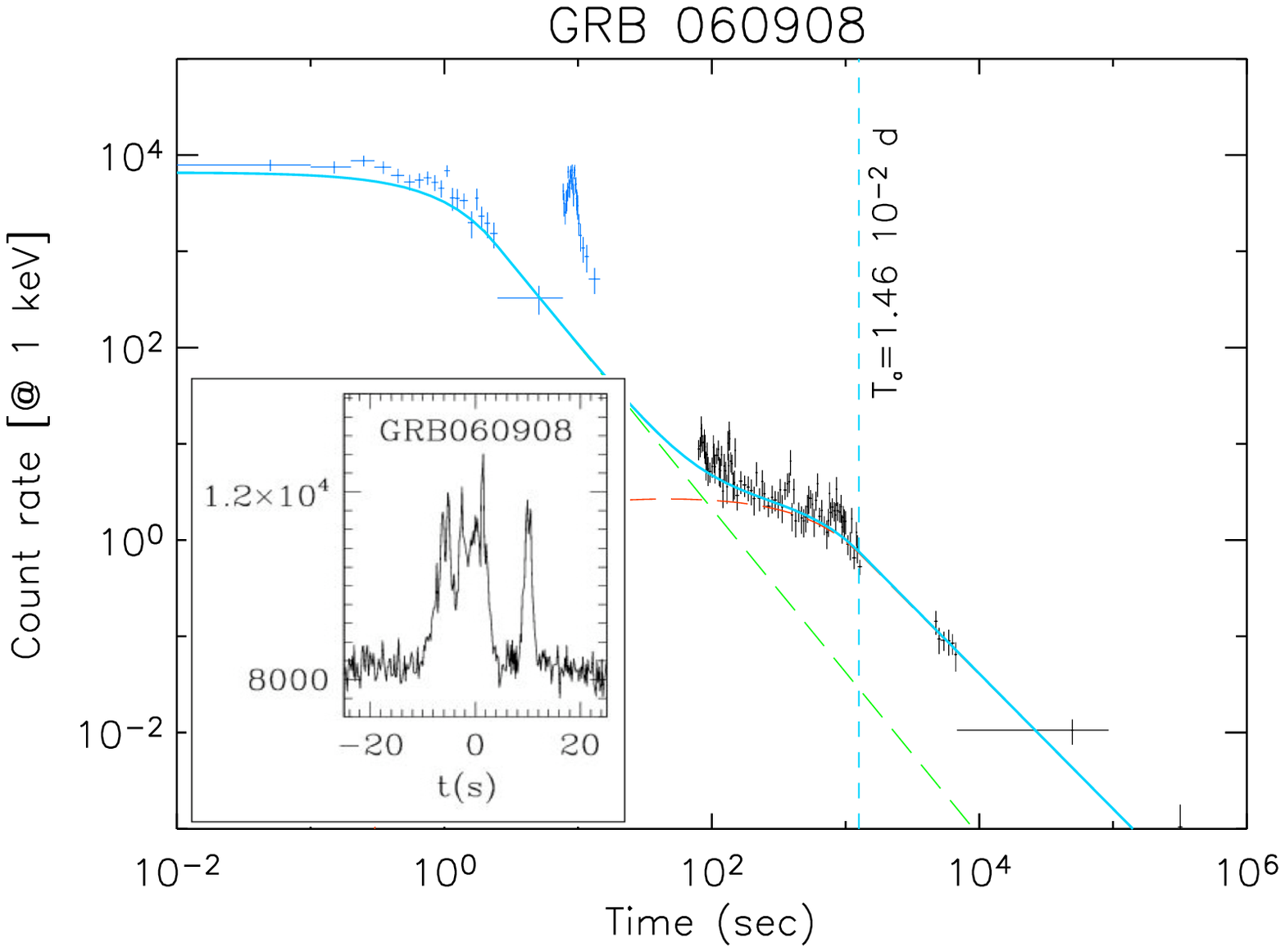,width=10.0cm}
  \vspace{-6cm}\\
  \hskip -1.5 cm
  \epsfig{figure=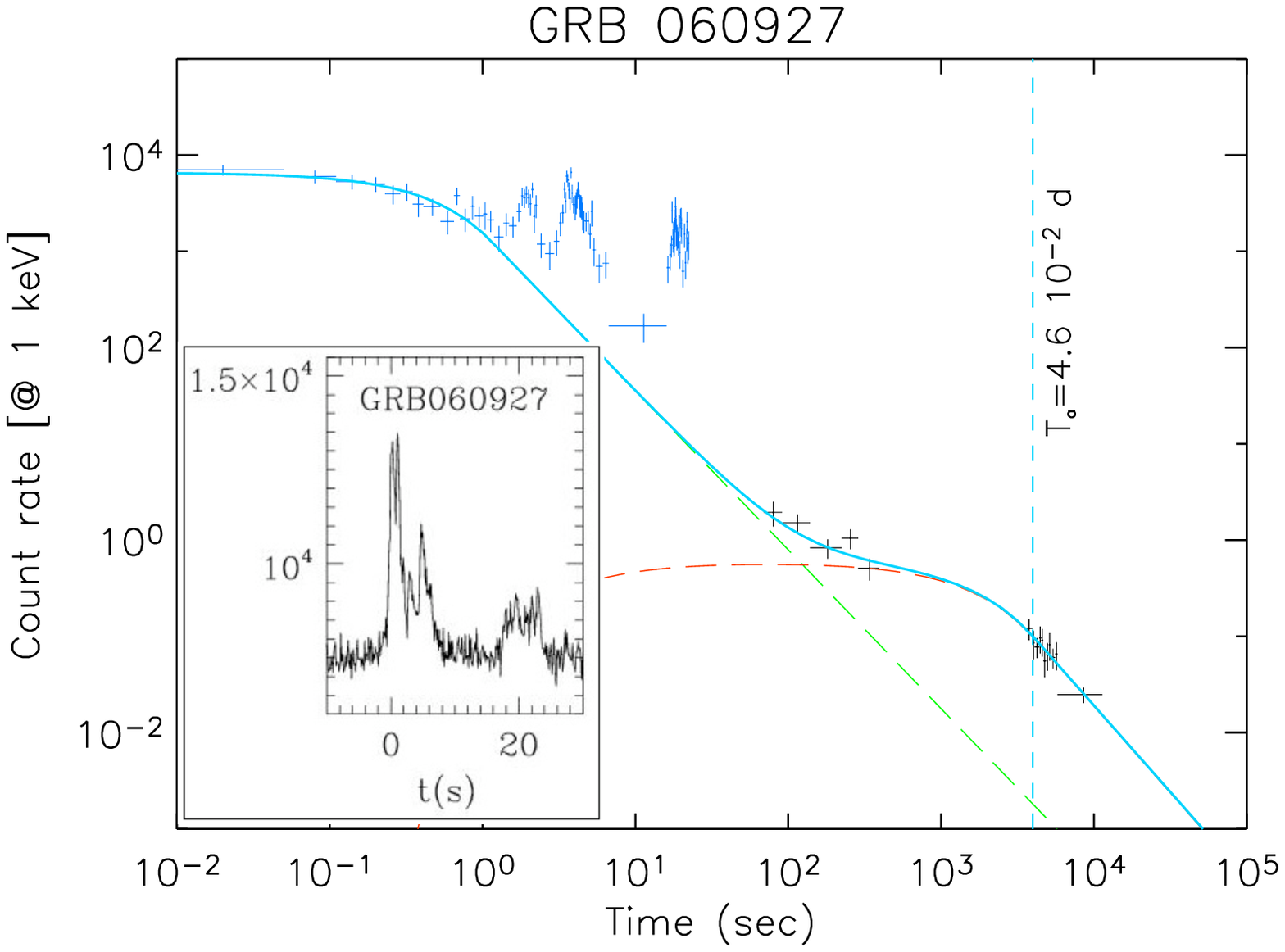,width=10.0cm}
   \hskip -1.5 cm
  \epsfig{figure=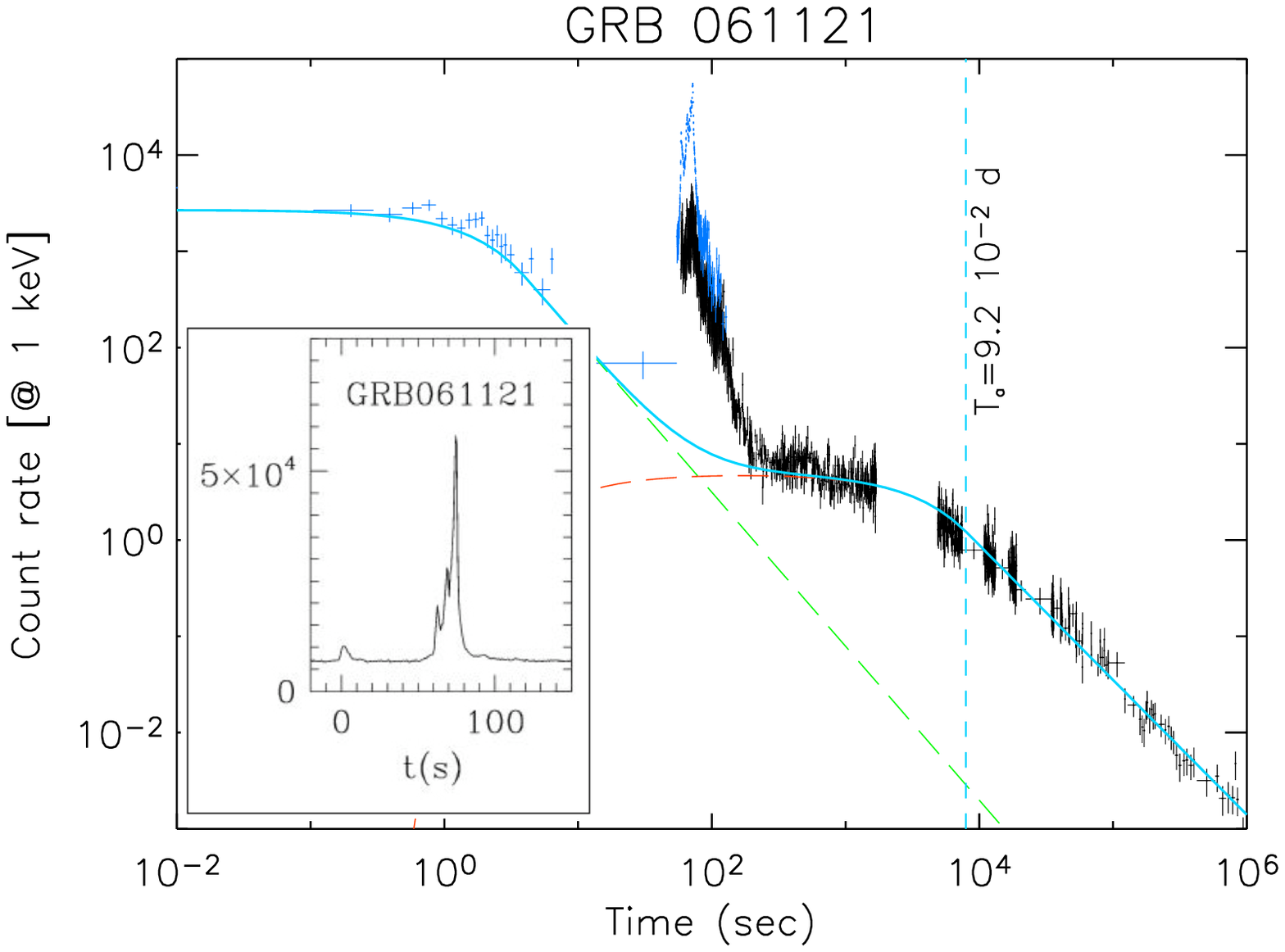,width=10.0cm}
\end{tabular}
\vspace{-5cm}
\caption{The XRT and (extrapolated) BAT light curves of the X-ray
emission from  GRBs 060124,   060908,   060927 and  061121 as
modeled by Nava et al.~(2007), following  Willingale et al.~(2006). The
two model components are shown by the long--dashed lines, their sum
is shown by the solid line. The fits were performed excluding the
X-ray flares. The long dashed vertical lines show the endtime of the
plateau phases. For GRB 060124,
the time of the alleged jet break in the optical light curve is also
shown. Inserted in each figure is the BAT $\gamma$-ray light-curve,
clearly showing that the early-time X-ray flares coincide in time with
the GRB pulses.}
\label{f2}
\end{figure*}

\newpage
\begin{figure}[]
\centering
\vspace{-2cm}
\vbox{
\hbox{
\epsfig{file=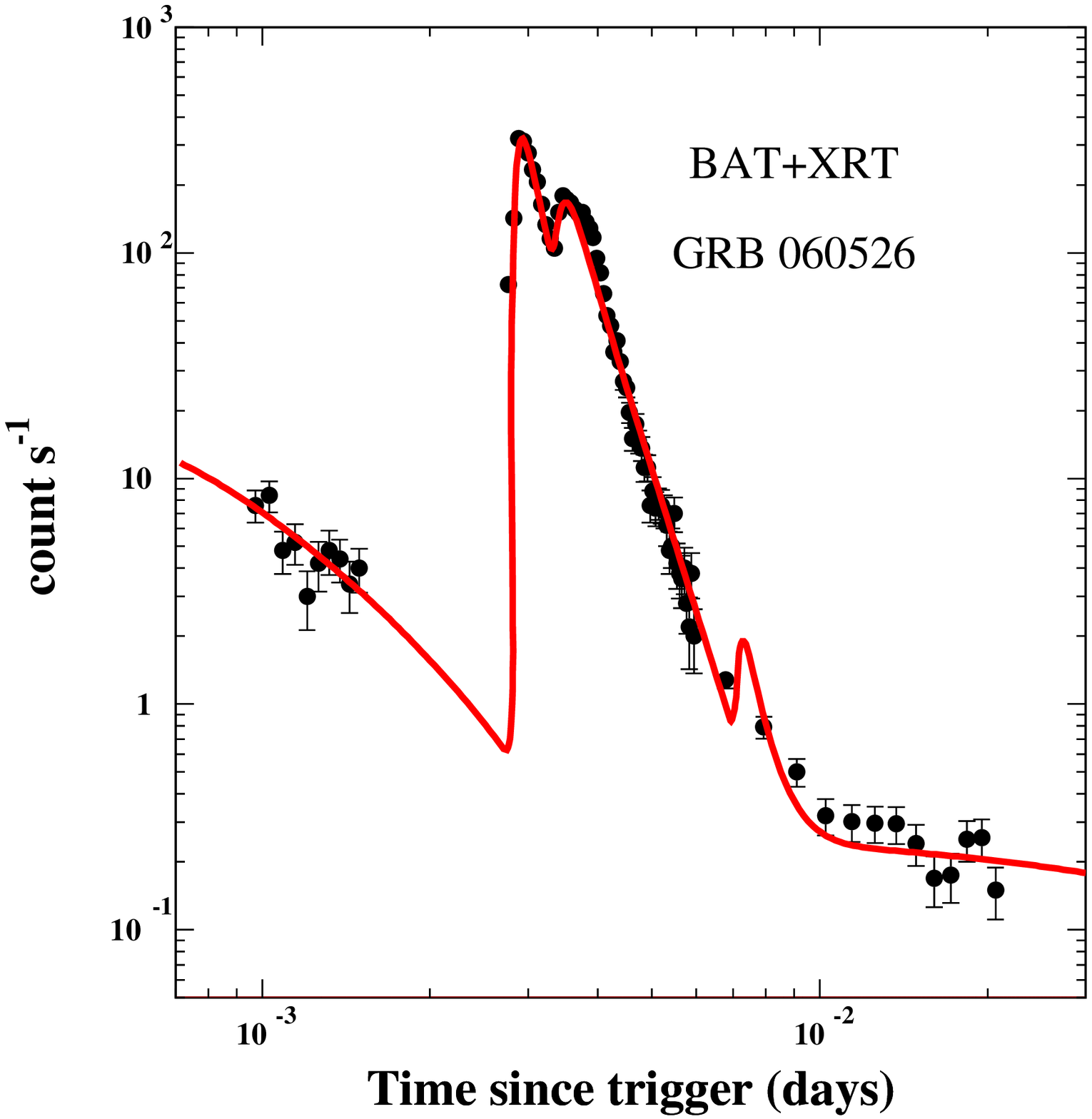,width=8.cm}
\epsfig{file=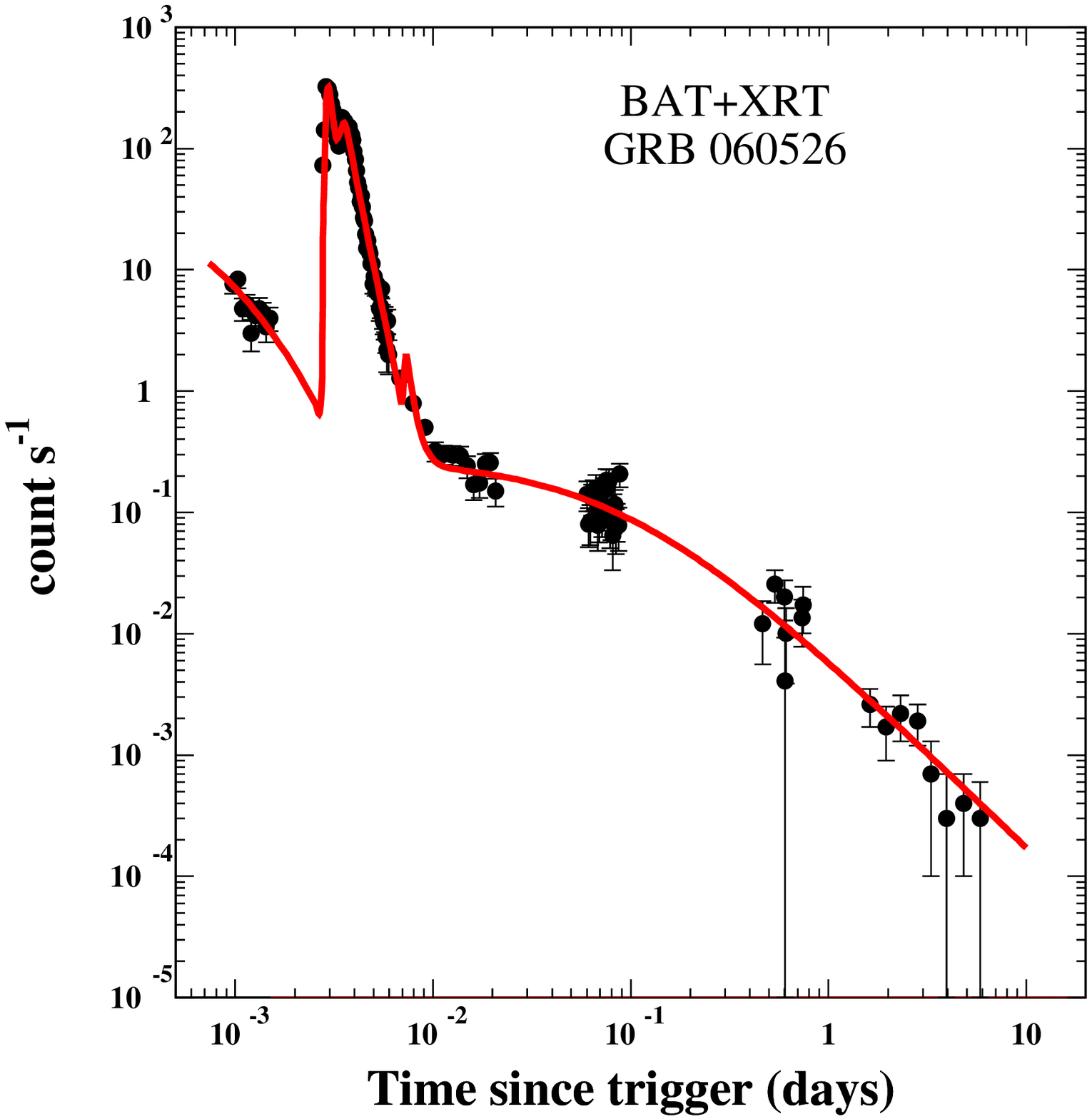,width=8.cm}
}}
\vbox{
\hbox{
\epsfig{file=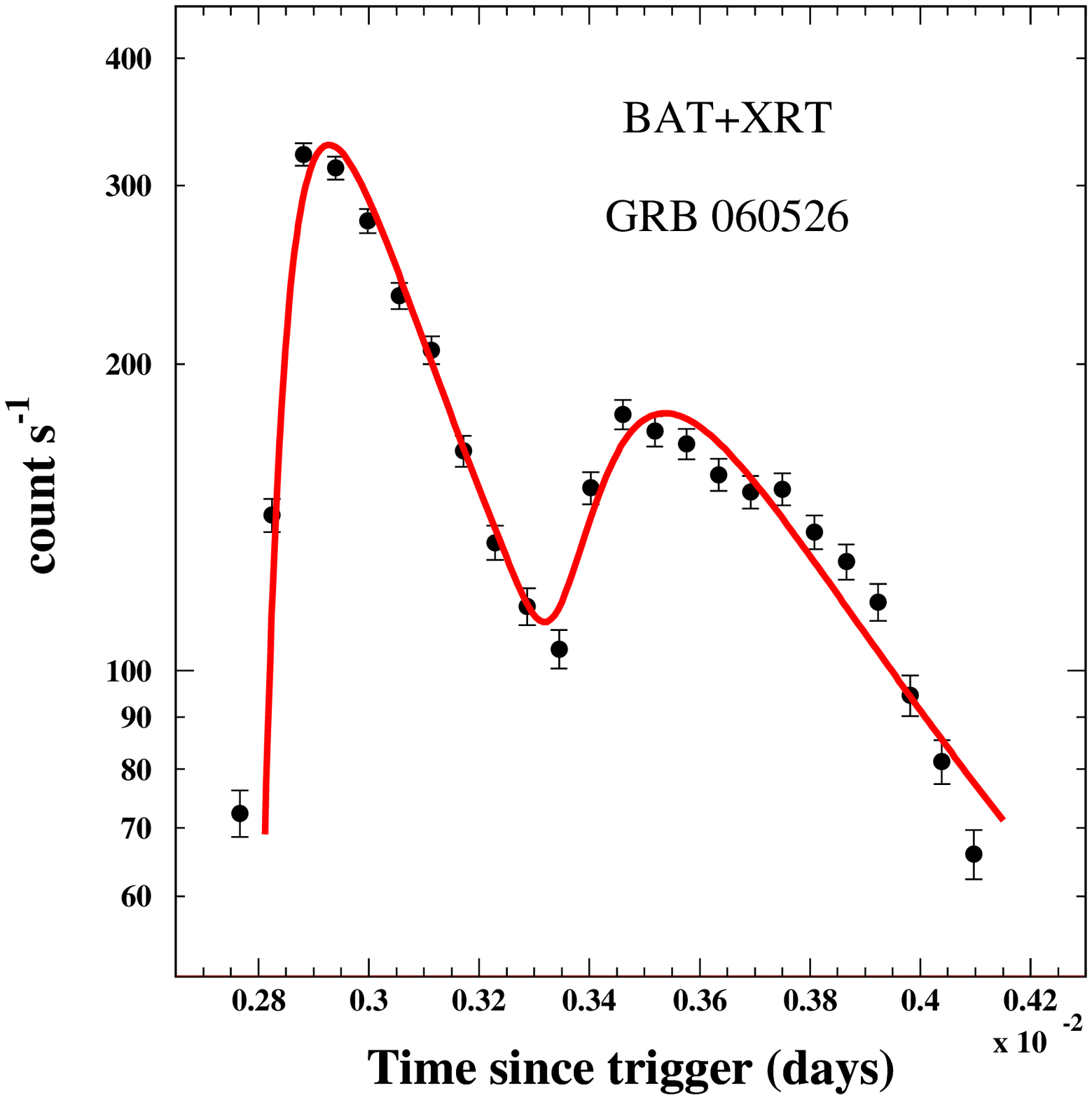,width=8.cm}
\epsfig{file=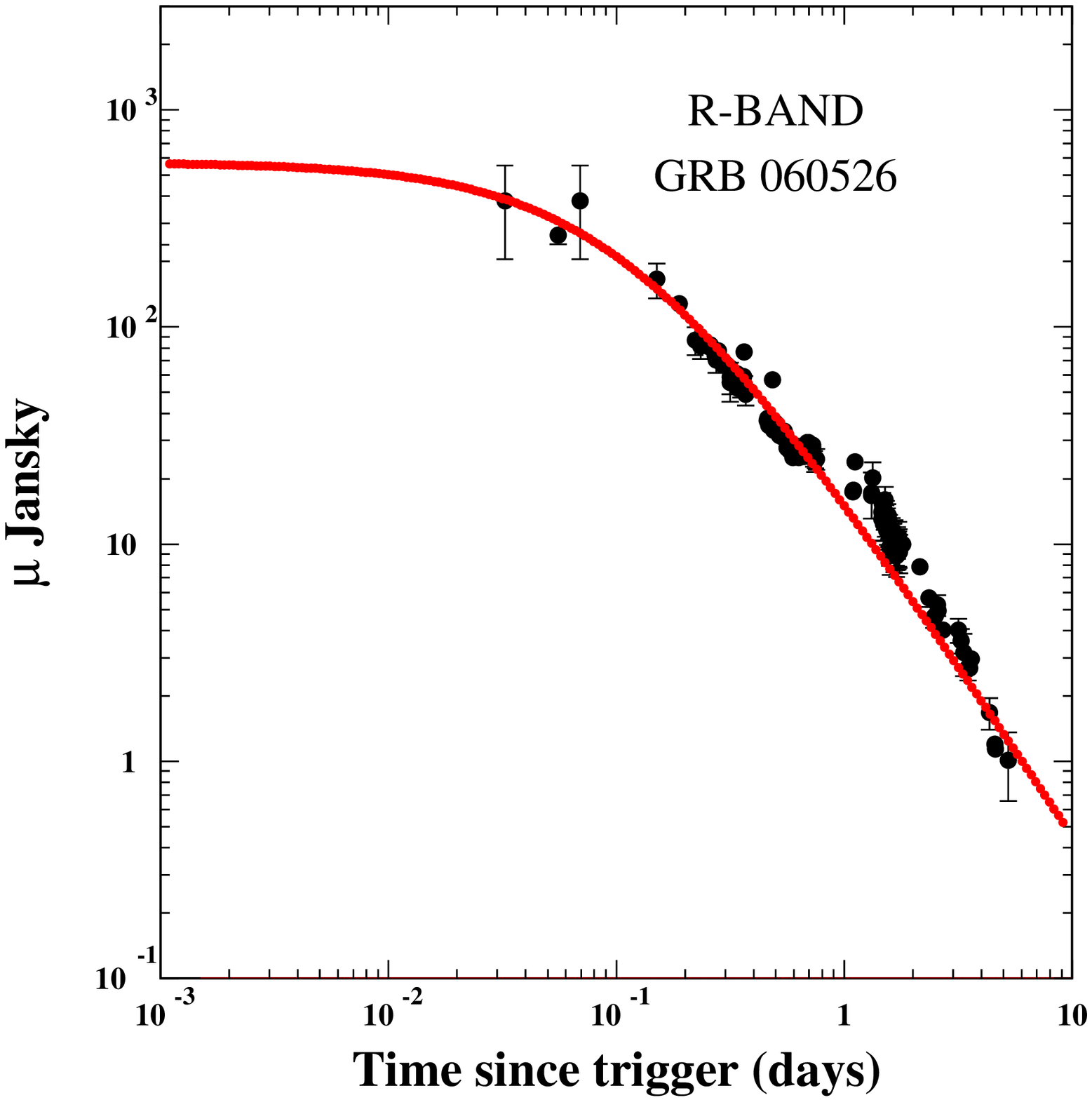,width=8.cm}
}}
\caption{Comparison of data  (Dai et al.~2007)
with the CB-model fit for GRB 
060526. {\bf Top left (a)}: The X-ray light curve as inferred from the SWIFT BAT and XRT data.
{\bf Top right (b):}
Enlarged view of the early-time X-ray light curve.
{\bf Bottom left (c):} Zoom-in on the major early-time flares in the
X-ray light curve.
{\bf Bottom  right (d):}
The R-band light curve (Dai et al.~2007 
and references therein). This curve is approximately achromatic
relative to the tail of the X-ray light curve in {\bf (a)}.}
\label{f3}
\end{figure}

\newpage
\begin{figure}[]
\centering
\vspace{-2cm}
\vbox{
\hbox{
\epsfig{file=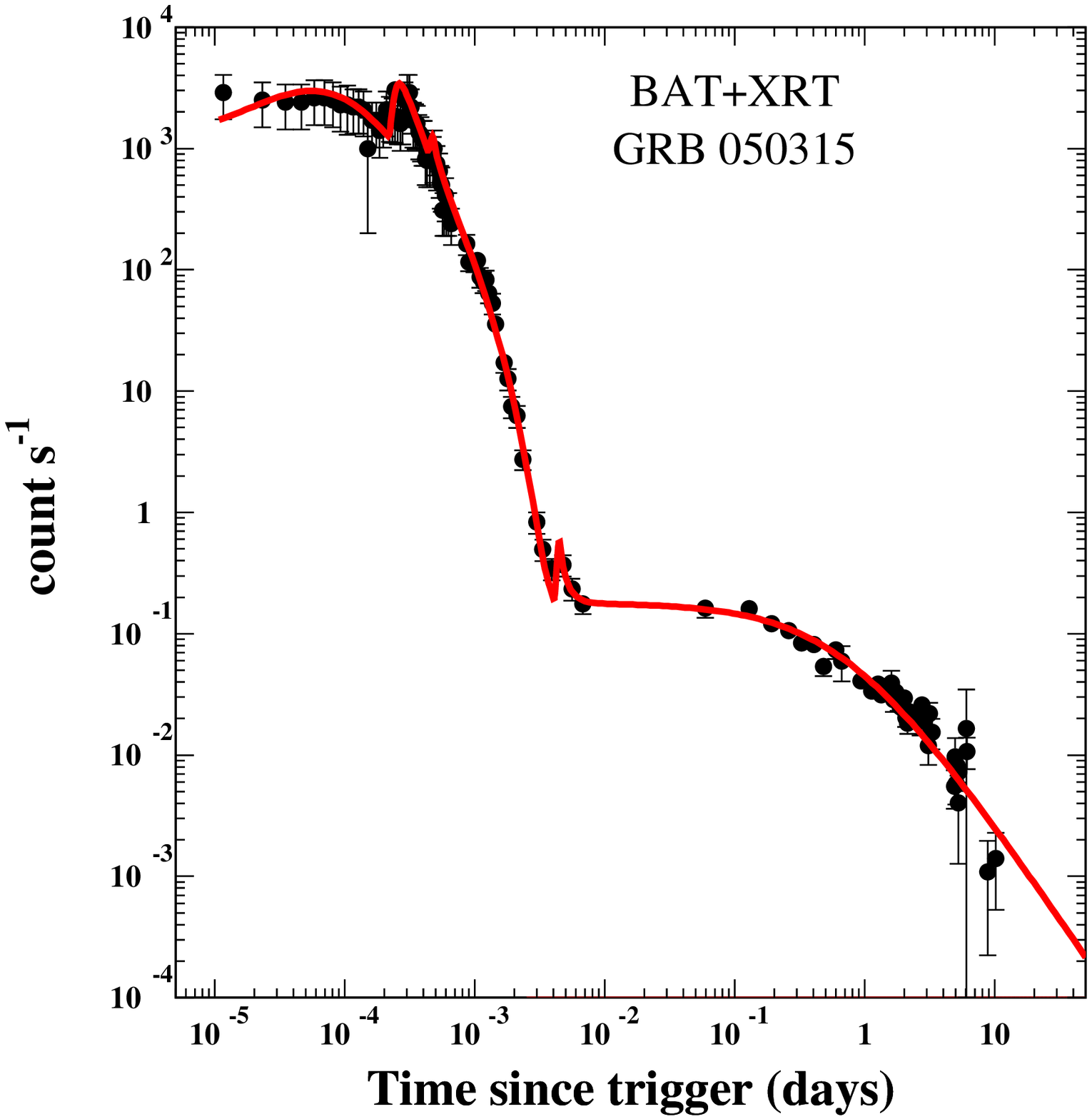,width=8.cm}
 \epsfig{file=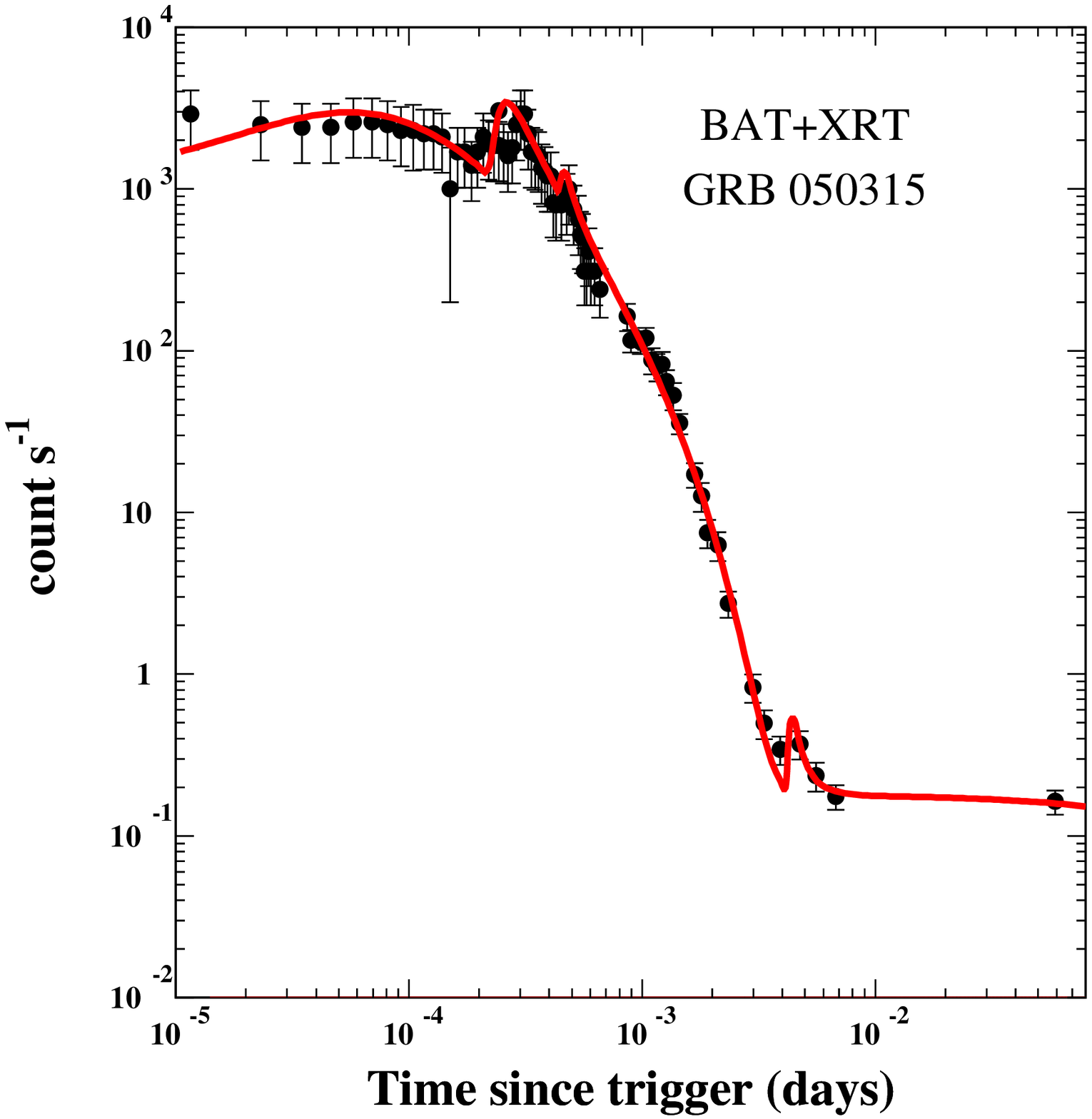,width=8.cm}
}}
\vbox{
\hbox{
 \epsfig{file=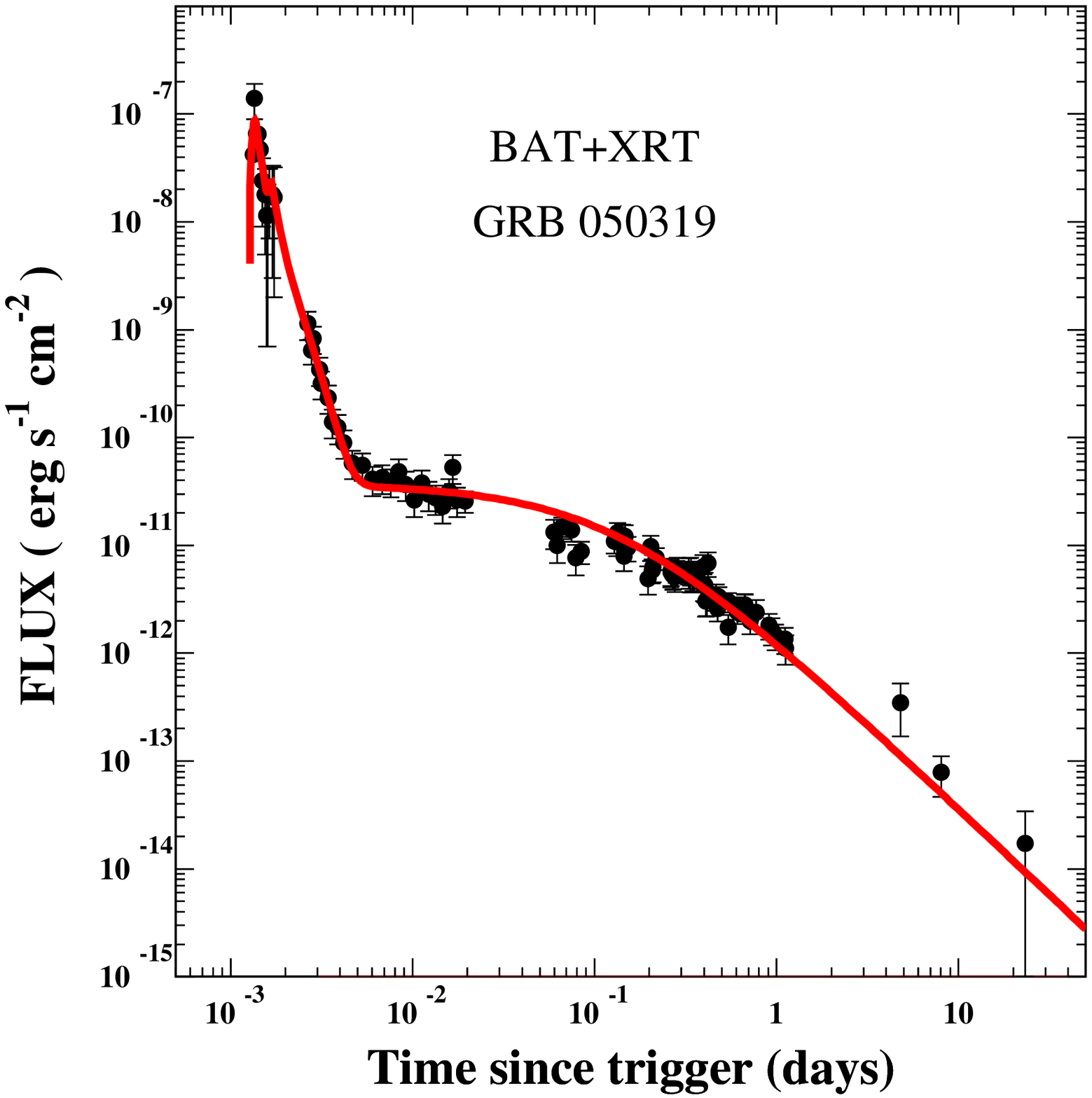,width=8.cm}
 \epsfig{file=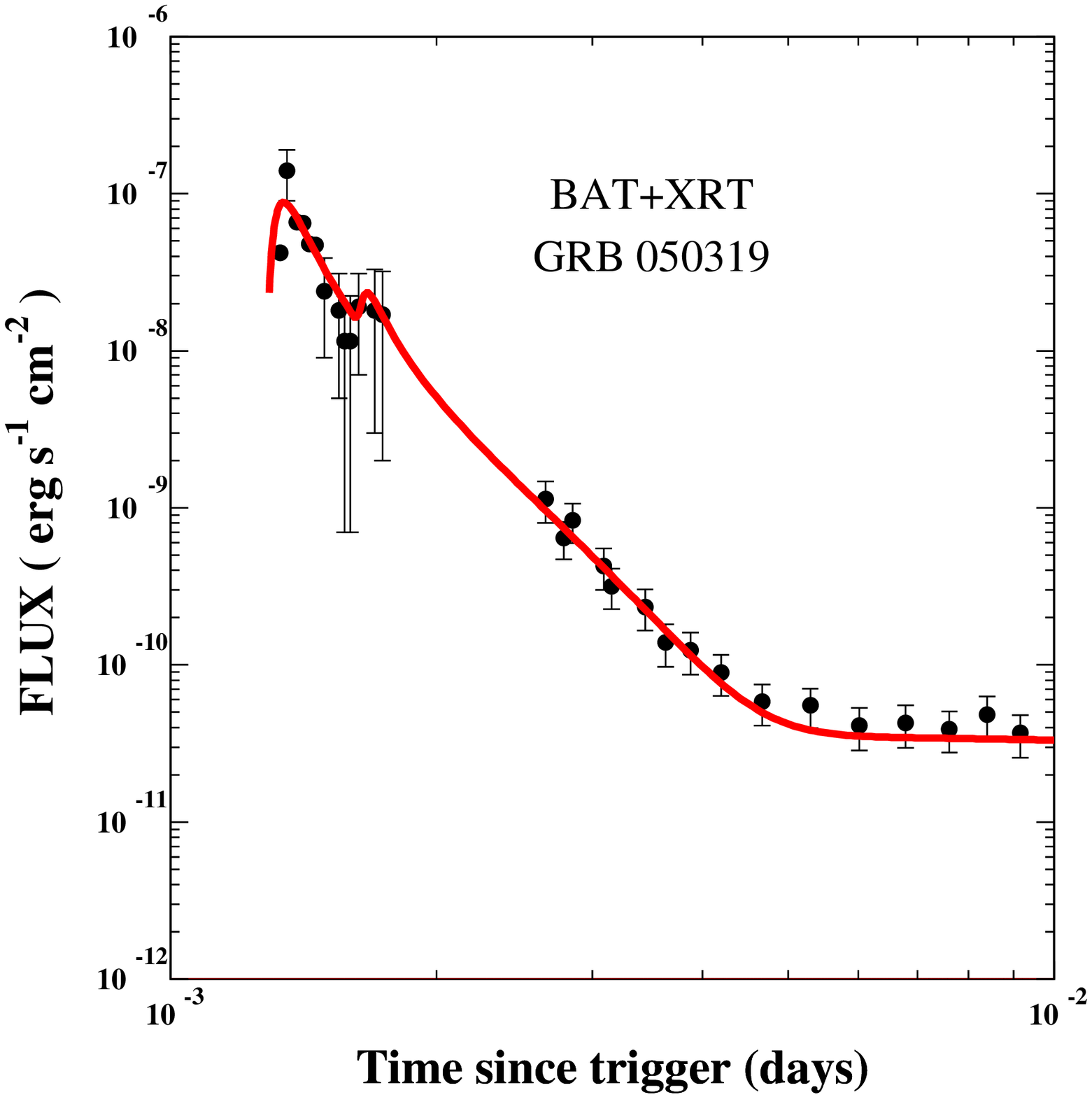,width=8.cm}
}}
\caption{  
Comparison of the CB model fits with the X-ray light curves
inferred from  SWIFT BAT and XRT data. {\bf Top:} GRB 050315
(Vaughan et al.~2006). {\bf Top left (a):} The data ensemble.
{\bf Top right (b):} Zoom onto the early data. {\bf Bottom:}
GRB 050319 (Cusumano et al.~2006).
{\bf Bottom left (c):} The data ensemble.
{\bf Bottom right (d):} Zoom onto the early data.}
\label{f4}
\end{figure}

\newpage
\begin{figure}[]
\centering
\vspace{-2cm}
\vbox{
\hbox{
\epsfig{file=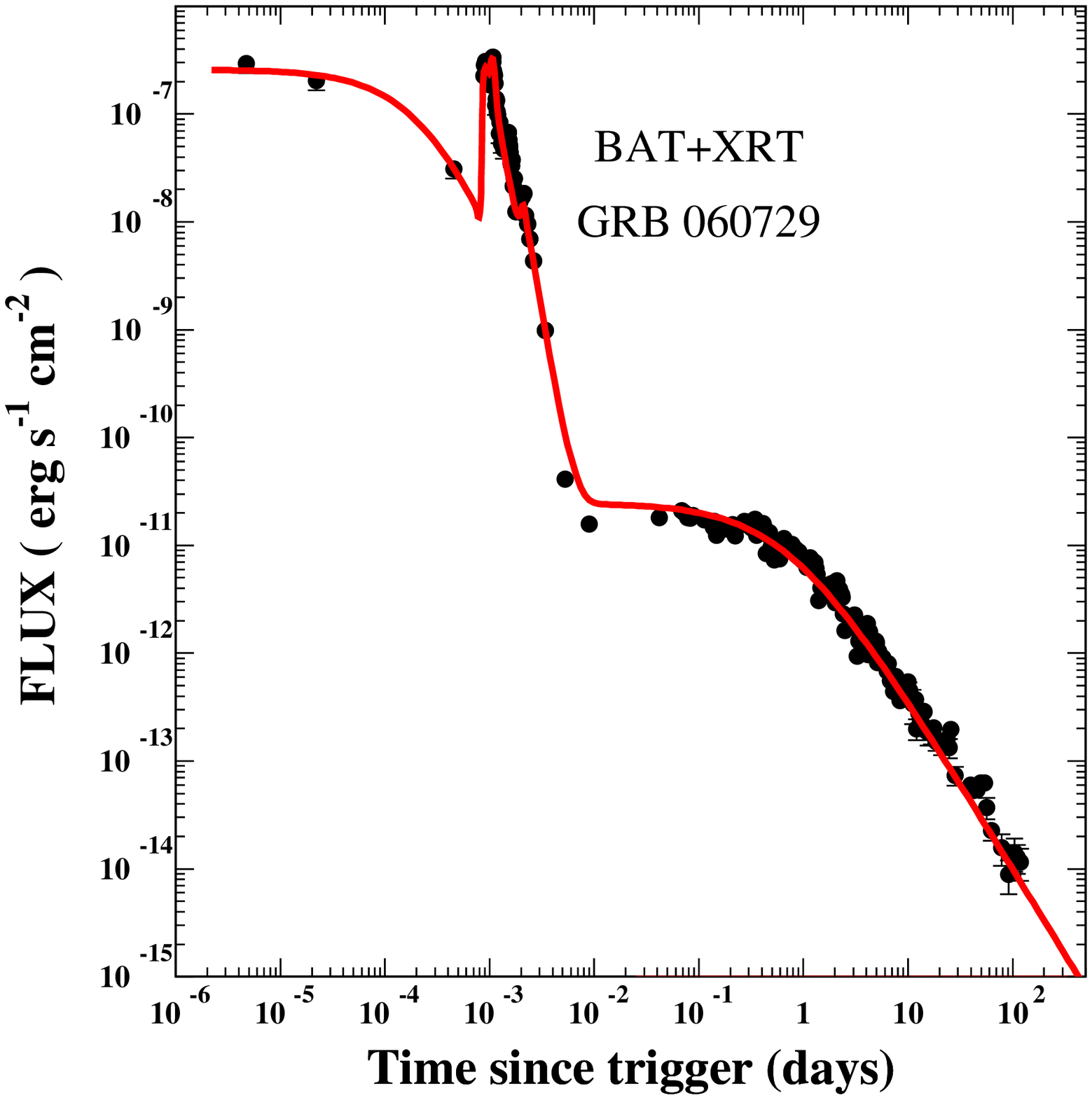,width=8.cm}
\epsfig{file=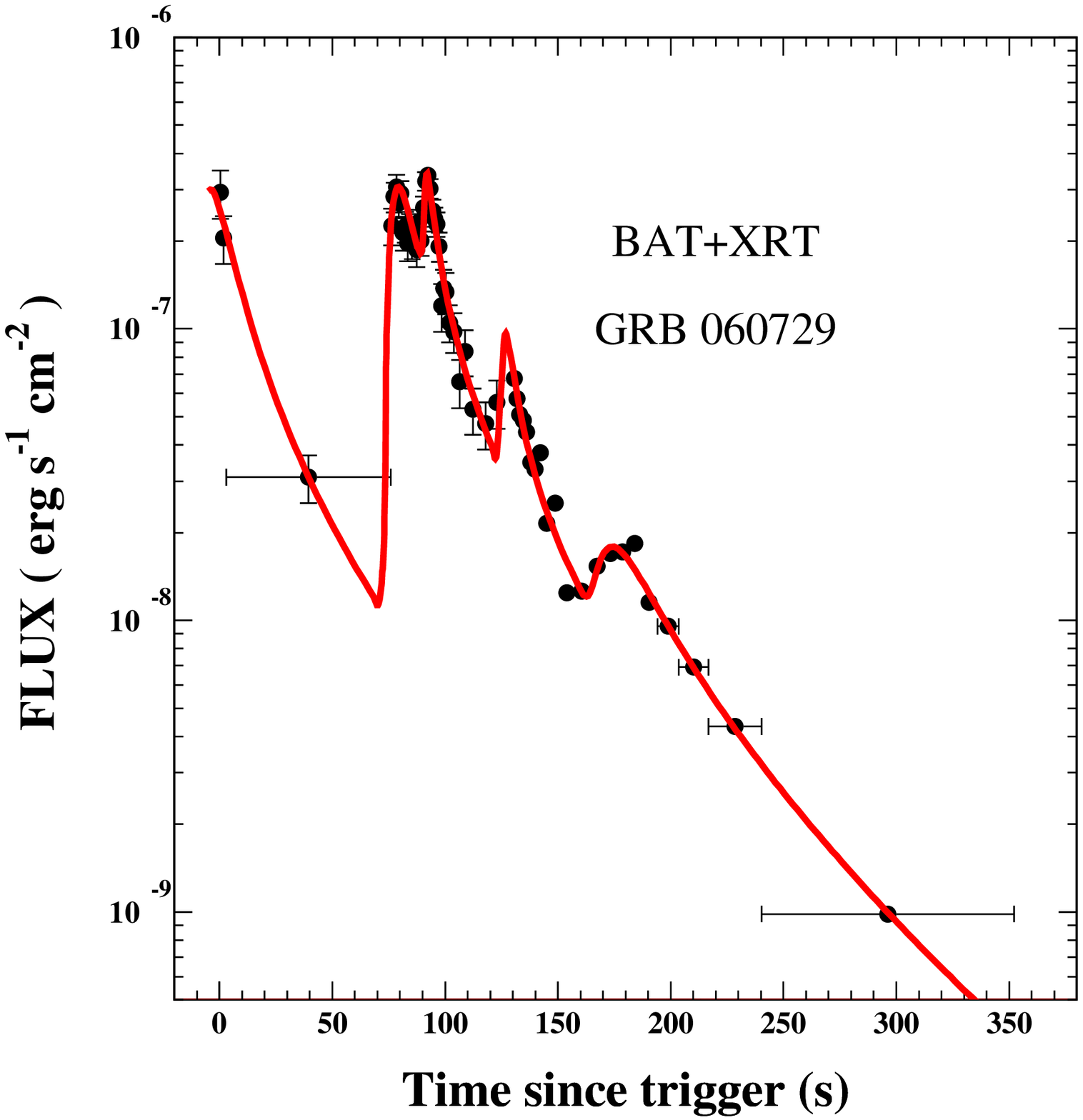,width=8.cm}
}}
\vbox{
\hbox{
\epsfig{file=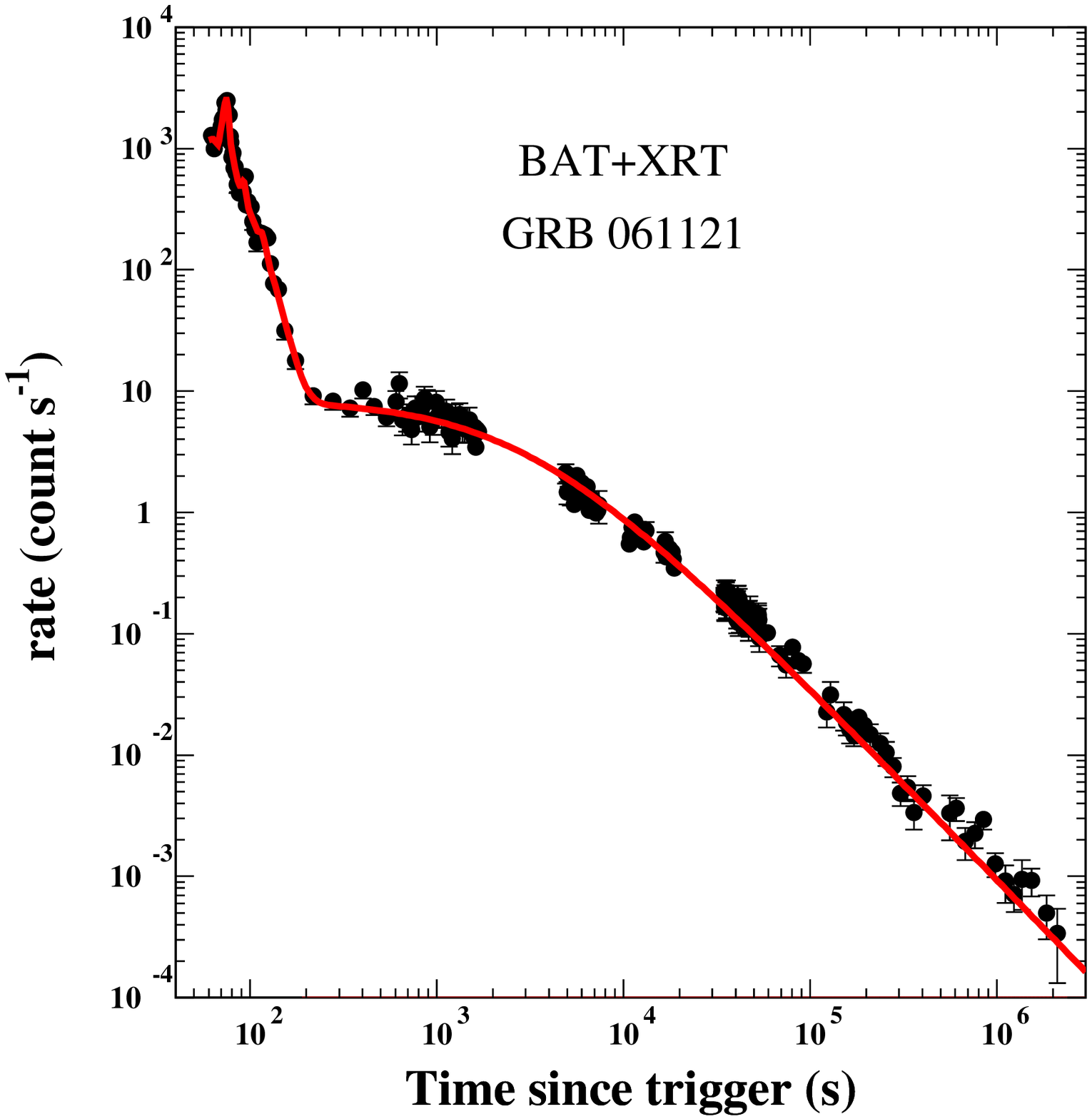,width=8.cm}
\epsfig{file=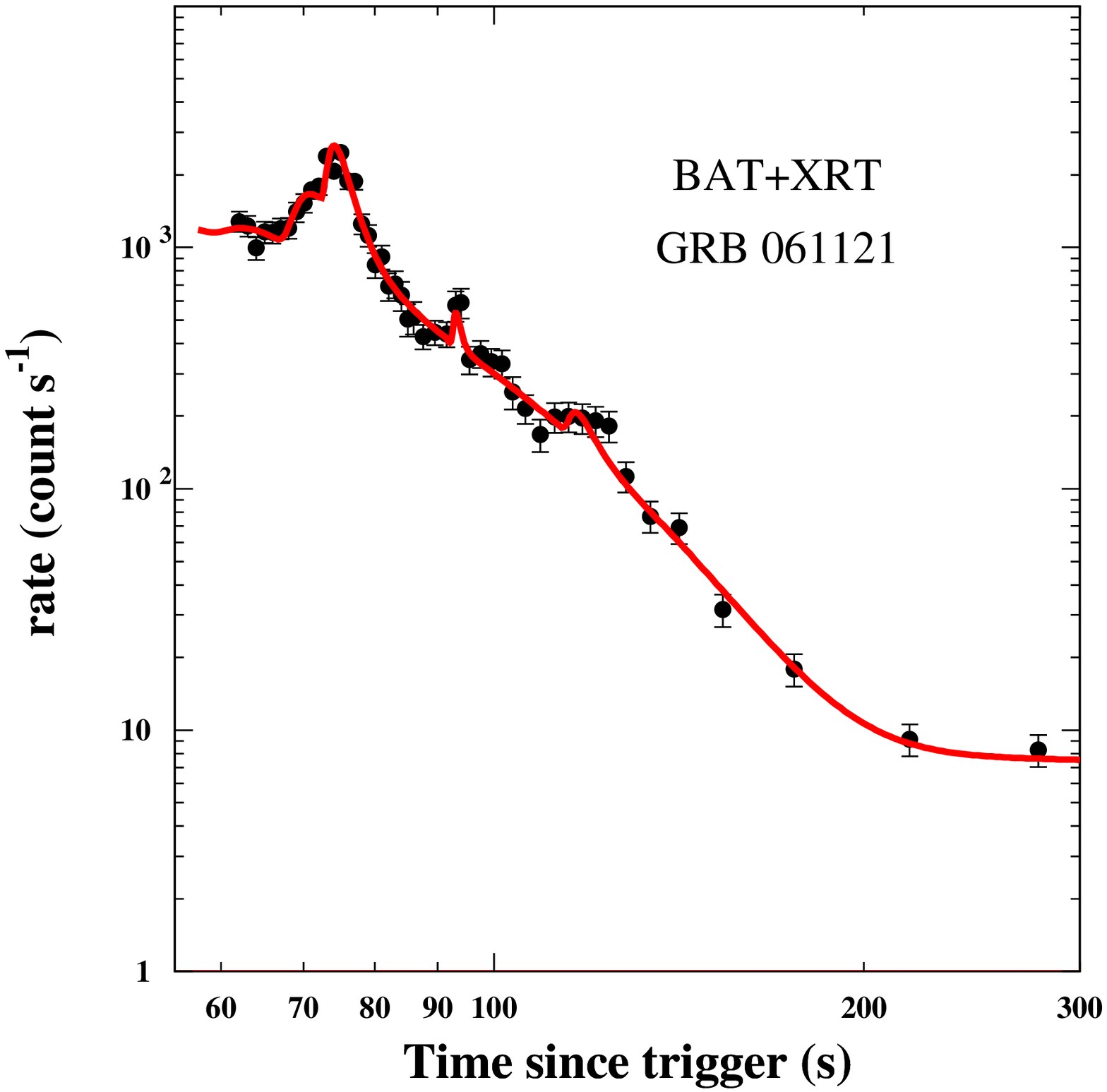,width=8.cm}
}}

\caption{  
Comparison of the CB model fits with the X-ray light curves
inferred from  SWIFT BAT and XRT data. {\bf Top:} GRB 060729
(Grupe et al.~2006), {\bf Top left (a):} The data ensemble.
{\bf Top right (b):} Zoom onto the early data. {\bf Bottom:}
GRB 061121 (Evans et al.~2006).
{\bf Bottom left (c):} The data ensemble.
{\bf Bottom right (d):} Zoom onto the early data.}
\label{f5}
\end{figure}

\newpage
\begin{figure}[]
\centering
\vspace{-1cm}
\vbox{
\hbox{
\epsfig{file=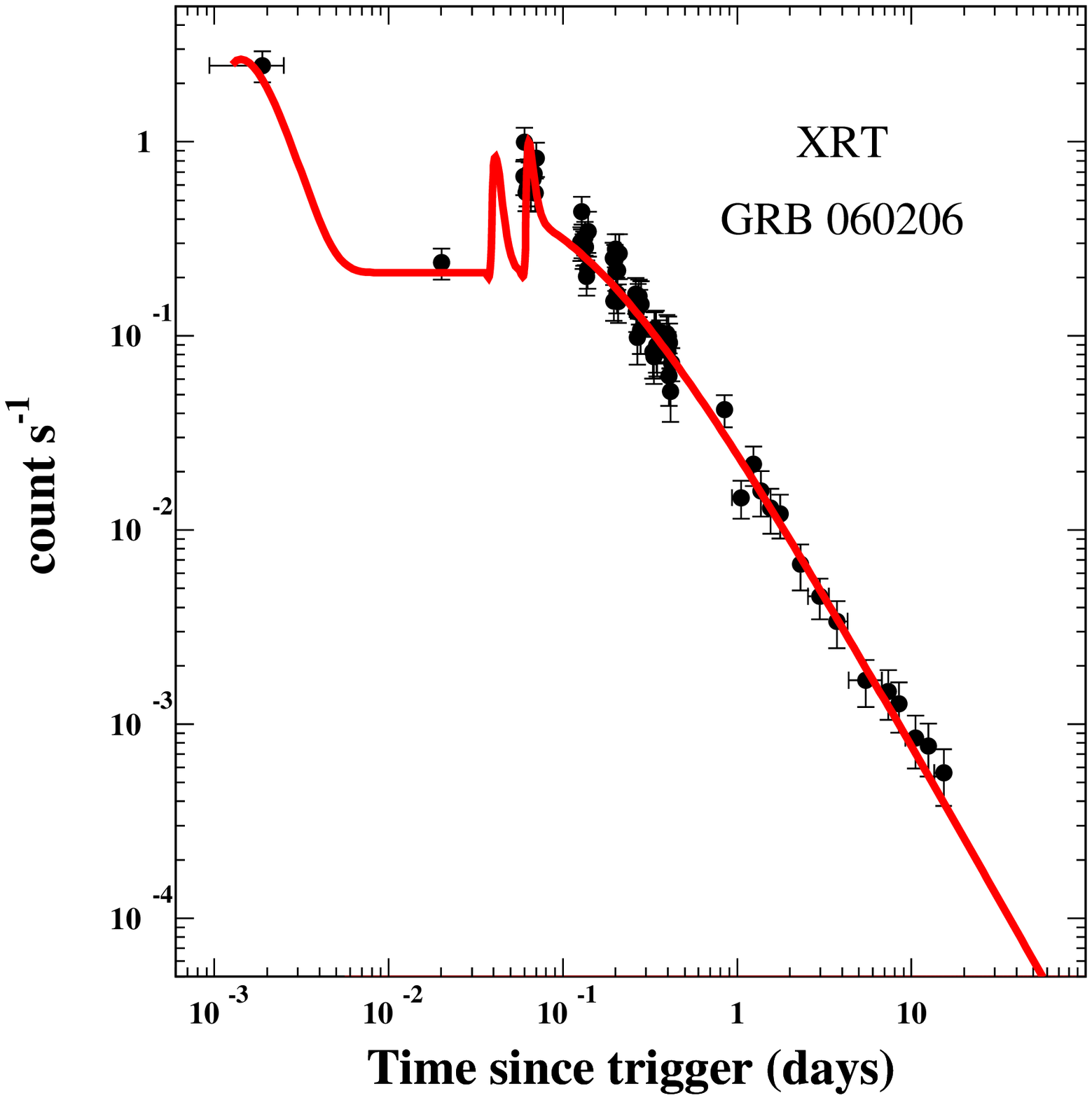,width=8.cm}
\epsfig{file=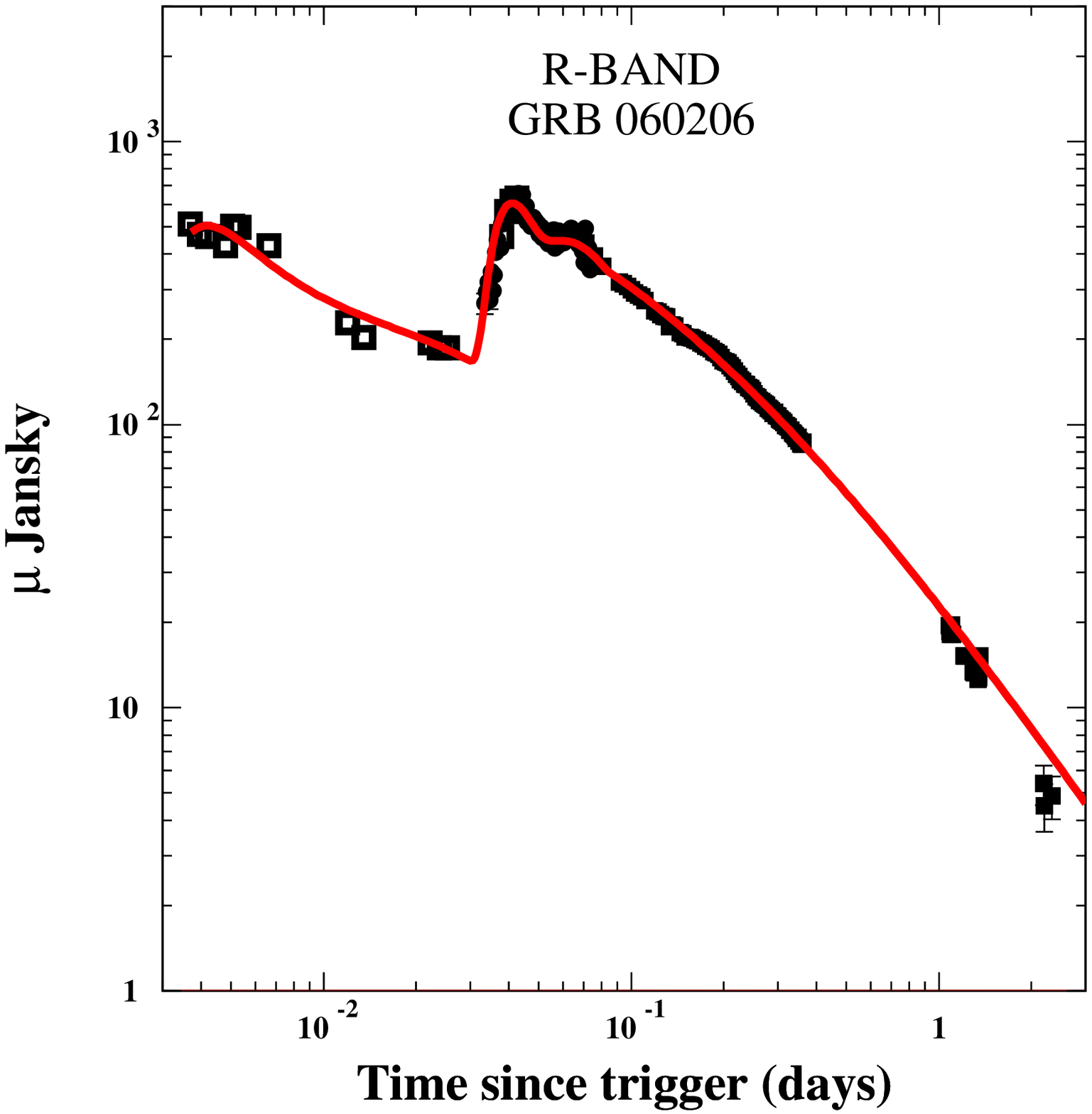,width=8.cm}
}}
\vbox{
\hbox{
\hskip 3.cm
 \epsfig{file=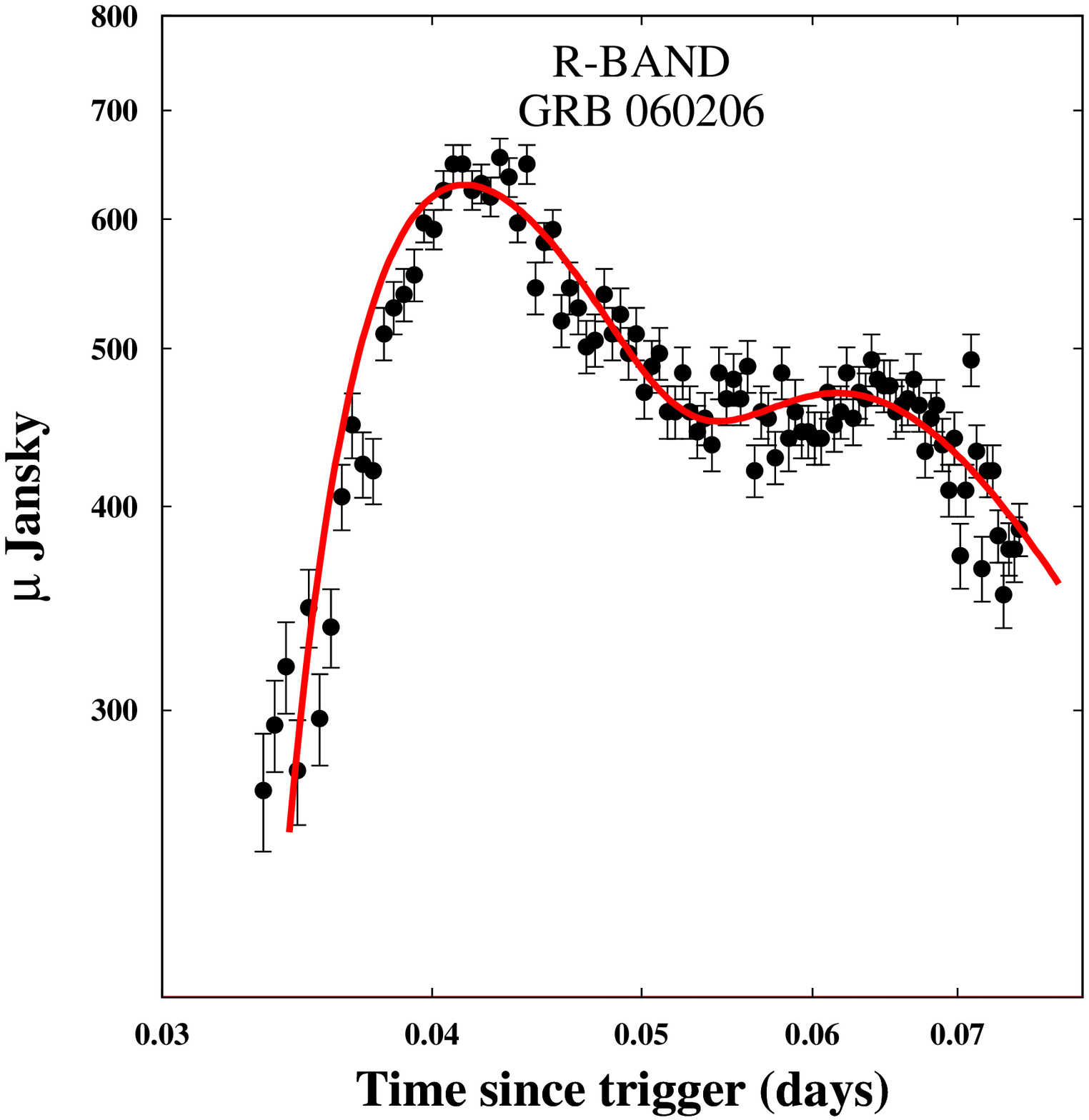,width=10.cm}
}}
\caption{  
Comparison of the CB model fits with the X-ray light curves 
(Evans et al.~2007)
inferred from  SWIFT BAT and XRT data and with the
R-band light curves (Wozniak et al.~2006, Stanek et al.~2007
  and Monfardini et al.~2006)  for GRB 060206.
 The parameters $\gamma_0$ and $\theta$ used in the X-ray
 prediction are extracted from the CB model fit
to the R-band light curve. The X-ray widths follow
from the `$E\!\times\!t^2$ law' and the observed
widths of the optical peaks.
{\bf Top left (a):} The entire X-ray light curve. 
{\bf Top right (b):} The entire R-band light curve. 
{\bf Bottom (c):} Enlarged view of the two major flares in the R-band
light curve 
(Wozniak et al.~2006). Note the linear time scale.
}
\label{f6}
\end{figure}

\newpage
\begin{figure}[]
\centering
\vbox{
\epsfig{file=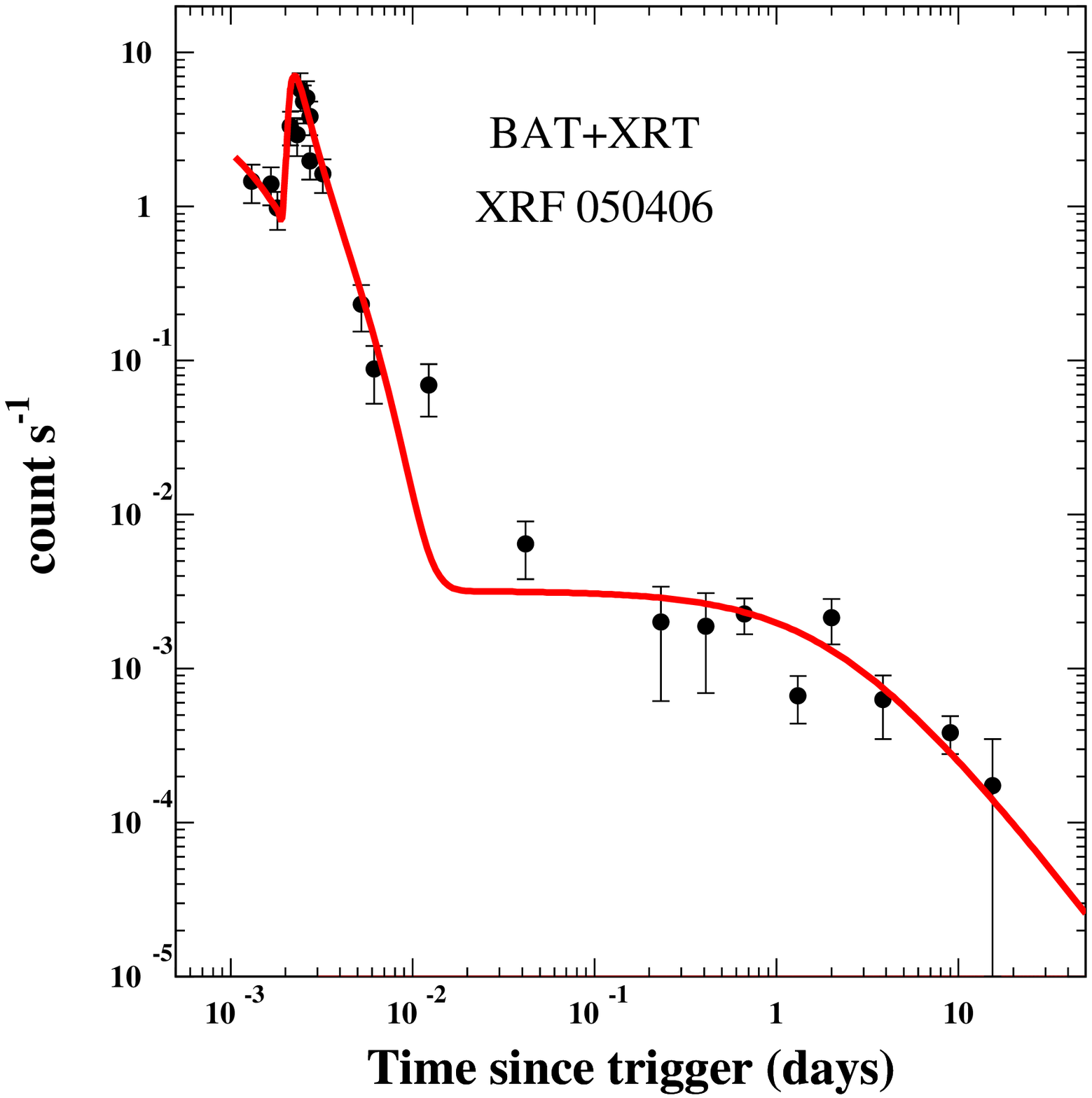,width=8.cm}
\epsfig{file=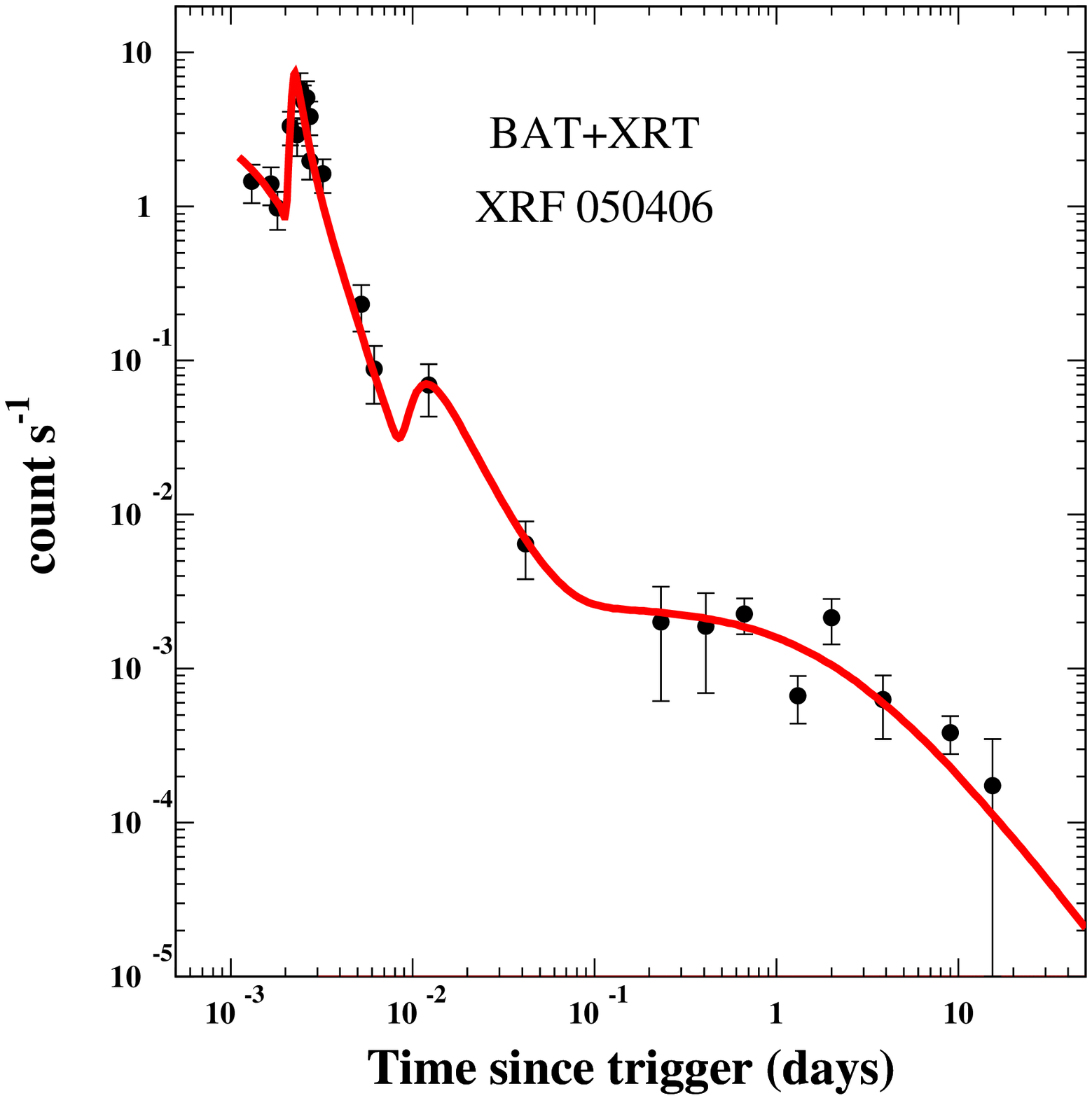,width=8.cm}
\hbox{
}}
\vbox{
\hbox{
\hskip 2.cm
\epsfig{file=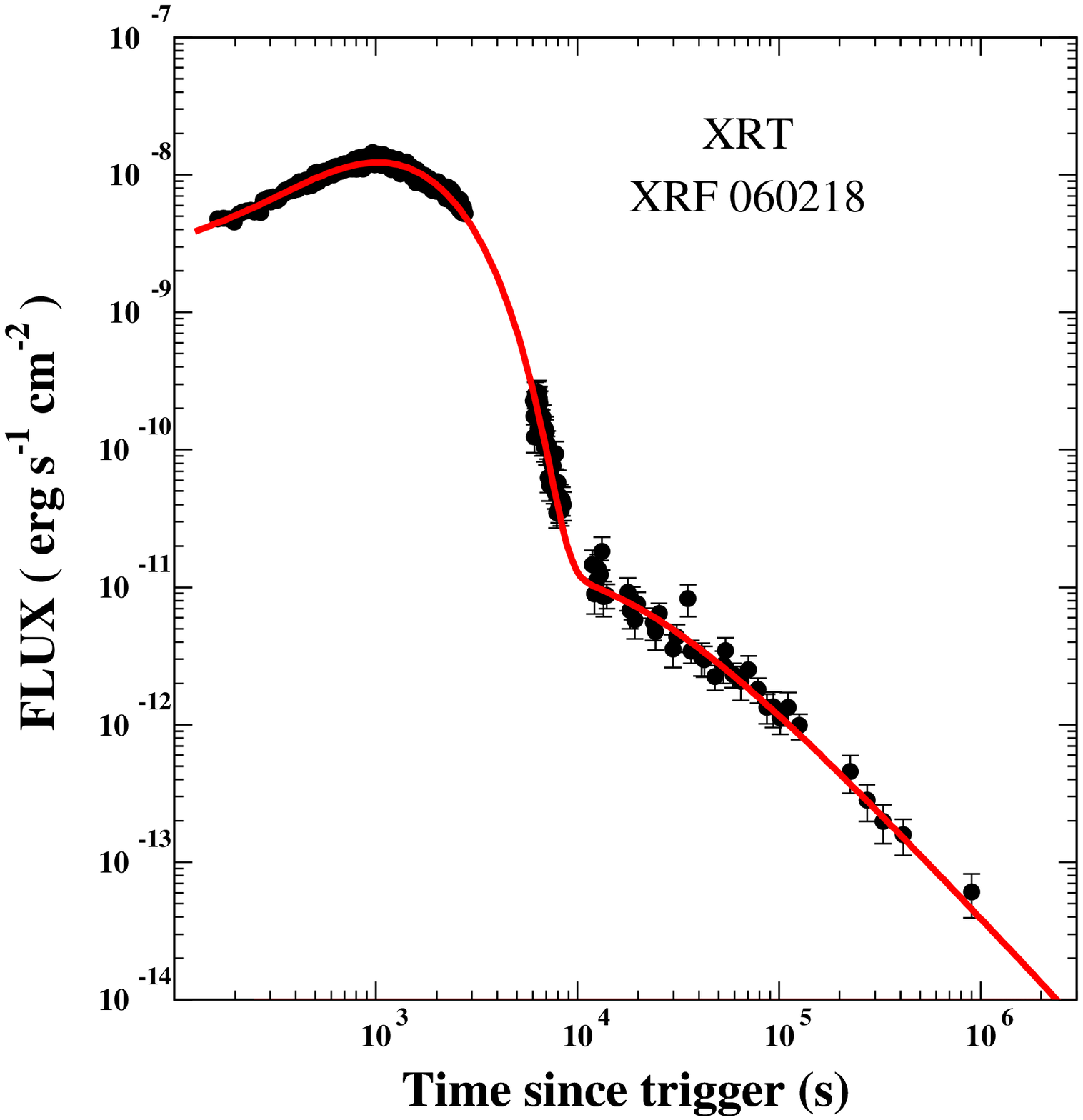,width=12.cm}
}}

\caption{  
Comparison of the CB model fits with the X-ray light curves 
inferred from  SWIFT BAT and XRT data  for XRF 050406
(Top, data from Romano et al.~2006) and for XRF060218
(Bottom, data from Campana et al.~2006b).
{\bf Top left (a):} The entire X-ray light curve, described
with two flares. 
{\bf Top right (b):} The entire X-ray light curve, described
with three flares. 
{\bf Bottom (c):} The entire X-ray data,
which has only one flare.
}

\label{f7}
\end{figure}

\newpage
\begin{figure*}
\begin{tabular}{cc}
  \hskip 2.5 cm
 \psfig{figure=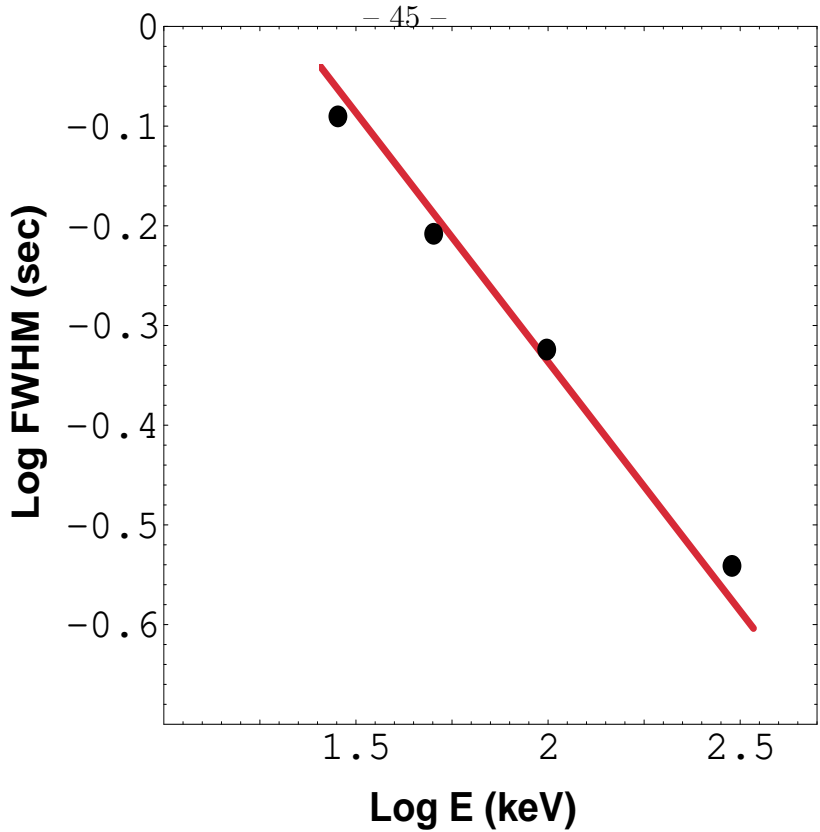,width=10.0cm,height=10.0cm}\\
  \hskip 3. cm
     \psfig{figure=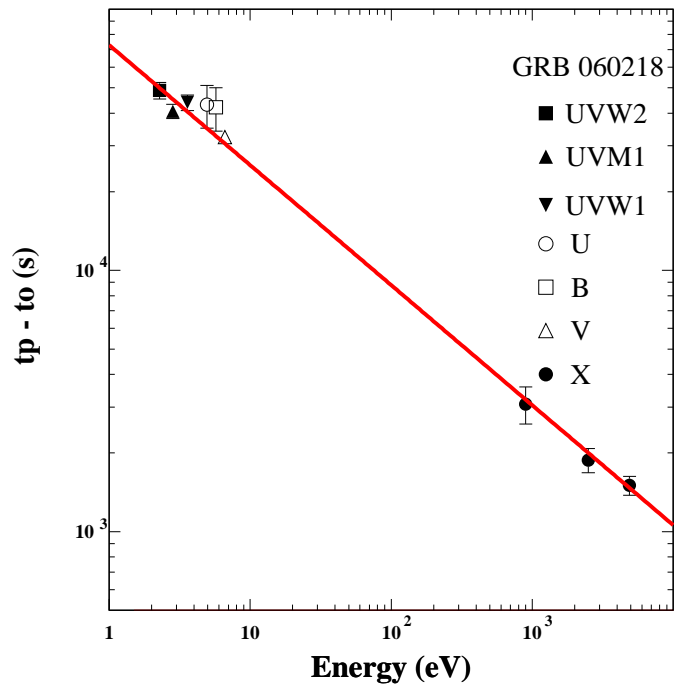,width=10.0cm,height=10.cm}
  \end{tabular}
\caption{{\bf Top:} Comparison between the  CB-model's prediction 
${\rm FWHM}\!\propto\! E^{-1/2}$ (DD2004) for the
average FWHM of GRB pulses with the
BATSE/CGRO data in its  
four energy channels
(Fenimore et al.~1995; Norris et al.~1996). {\bf Bottom:}
Comparison between the same predicted dependence of
peak-time on energy, and the SWIFT UVOT and XRT
data for XRF060218 (Campana et al.~2006b; Dai et al.~2007).
Notice how well 
the originally-tested prediction (DD2004) extends to the optical 
domain, more than three orders of magnitude away in energy. 
The peak-energy fluxes are equally well predicted, see Table
\ref{t3}.}
\label{f8}
\end{figure*}

\end{document}